\documentclass[preprintnumbers,amsmath,amssymb,floatfix,11pt,prd,onecolumn,superscriptaddress,nofootinbib]{revtex4}
\usepackage{diagbox}
\usepackage{latexsym}
\usepackage{epsfig}
\usepackage{epstopdf,float}
\usepackage{graphicx}
\usepackage{amssymb}
\usepackage{amsmath}
\usepackage{dcolumn}
\usepackage{bm}
\usepackage{color}
\usepackage{comment}
\usepackage[shortlabels]{enumitem}
\usepackage{subfigure}
\usepackage{multirow}
\usepackage{xcolor}
\begin{document}
\title{\bf Probing Horndeski Gravity via Kerr Black Hole: Insights from Thin Accretion
Disks and Shadows with EHT Observations}
\author{Xiao-Xiong Zeng}
\altaffiliation{xxzengphysics@163.com}\affiliation{College of
Physics and Electronic Engineering, Chongqing Normal University,
Chongqing 401331, China}
\author{Chen-Yu Yang}
\altaffiliation{chenyu\_yang2024@163.com}\affiliation{Department of
Mechanics, Chongqing Jiaotong University, Chongqing $400074$, China}
\author{Muhammad Israr Aslam}
\altaffiliation{mrisraraslam@gmail.com}\affiliation{Department of
Mathematics, COMSATS University Islamabad, Lahore Campus,
Lahore-$54000$ Pakistan.}
\author{Rabia Saleem}
\altaffiliation{rabiasaleem@cuilahore.edu.pk}\affiliation{Department
of Mathematics, COMSATS University Islamabad, Lahore Campus,
Lahore-$54000$ Pakistan.}

\begin{abstract}
In this study, we have considered the Kerr-like black hole (BH)
model in Horndeski gravity and analyse the visual characteristics of
shadow images under two illumination models, such as a celestial
light source and a thin accretion disk. To capture the BH shadow
images, we utilise a recent fisheye camera model and ray-tracing
procedures. In this view, we carefully addressed the influence of
the spin parameter $a$ and the hair parameter $h$ on the BH shadow
images. The results indicate that for smaller values of $h$, the BH
shadow contours shift noticeably towards the right side of the
screen, while for larger values of $h$, the nearly circular shadow
gradually deforms into a ``D'' shape profile. For a celestial light
source, the larger values of $h$ lead to a reduction in the
corresponding radius of the photon ring, while the space-dragging
effect becomes more pronounced with increasing $a$. We further
discuss the distinctive characteristics of images observed in both
prograde and retrograde accretion disk scenarios. The results reveal
that variations in $h$ significantly affect both the inner shadow
and the resulting Einstein ring. Subsequently, we also discussed the
distinct features of red-shift configurations on the disk for both
direct and lensed images, which are closely related to the accretion
flow and the relevant parameters. We also attempt to use the recent
observational data from M$87^{\ast}$ and Sgr $A^{\ast}$ and
constrain the hair parameter $h$, confirming the validity of
Horndeski gravity.
\end{abstract}
\date{\today}
\maketitle

\section{Introduction}
The expanding nature of our universe can be described as the study
of evolution, the vast number of experimental and productive changes
that have been completed during all time and across all space from
the big bang to humankind. In recent years, the Einstein's theory of
general relativity (GR) has prevailed as the most successful theory
in weak and strong gravitational fields \cite{en1}. Besides its
successful description, it needs to be further generalized to deal
with some important problems, such as an inflationary phase of the
early universe, the nature of the dark matter and dark energy, the
Hubble tension and the cosmological constant. In this scenario, over
the past two decades, scientists have dedicated significant
theoretical and observational efforts to deepen humanity's
understanding of this cosmic phenomenon. In addition GR encapsulates
singularities such as, primordial Big Bang singularity, which are
still puzzling for the scientific community, and need to be a
quantum description of gravity for a better understanding
\cite{en2}. To address these challenges, one can consider modified
theories of gravity, which offer a theoretical scenario for
understanding various physical phenomena occurring in the universe
\cite{en3}. Consequently, to properly understanding the nature of
gravity, it is necessary that we consider hypothesis beyond the GR.
Among many modified theories of gravity, scalar-tensor theories
\cite{en4,en5} indicate the most fundamental alternative theory of
gravity that involves a scalar field in addition to the metric
tensor $g_{\mu\nu}$ for gravitational interaction. In 1974,
Horndeski has developed the most well-known four-dimensional
scalar-tensor theories \cite{en6}, so-called Horndeski gravity, by
inspiration of the contribution of Lovelock \cite{en7}.

Horndeski theory involve additionally only a single scalar degree of
freedom, which lead to produce second-order field equations, and are
free of ghosts, which makes them more attractive. This theory can be
regarded as a unifying framework, encompassing several well-known
theories such as, GR, Brans-Dicke theory, $f(R)$ gravity (where $R$
is a Ricci scalar), and quintessence models etc. In this regard, it
serves as one of the most significant alternatives of GR. In the
framework of Horndeski gravity, several significant properties of BH
solutions are investigated through effective implementations such as
$P-V$ criticality \cite{en8}, BH thermodynamics \cite{en9,en10},
holographic applications \cite{en11,en12,en13}, shadows of BH under
different accretions \cite{en14} and other attractive features
\cite{en15} etc. Moreover, the space-time of Horndeski gravity is
discussed with the hairy BH in \cite{en16,en17}. In literature,
during the last decades, many static and spherically symmetric hairy
BH solutions were derived in scalar-tensor theories. The simplest
case admitting solutions with a hairy profile characterized by a
radially dependent scalar field was studied in
\cite{en18,en19,en20}. The case in which the scalar field linearly
depends on time was discussed in \cite{en21,en22}. Indeed, the
authors in \cite{en23} derived the time-dependent hairy BH solutions
within the framework of Horndeski gravity. In the context of
Horndeski gravity, the exact static spherically symmetric hairy BHs
solutions are derived in \cite{en24}. By assuming the spherically
thin accretion flow matter, the authors in \cite{en25} investigated
the optical appearance of static hairy Horndeski BHs for different
values of involved parameters. By considering the steady-state
Novikov-Thorne model, the optical signatures of thin accretion disk
for rotating hairy BHs are investigated in \cite{en26}.
Particularly, they consider the background of the Horndeski gravity
and explore the impact of hair parameter on both the disk properties
and its image.

In recent decades, the development of experimental advances in the
field of BH physics have opened up a new avenue to investigate their
shadows and its related intricate properties. The detection of first
gravitational signals, which had emanated from the merger of two BHs
by LIGO-Virgo collaboration \cite{sd1,sd2}, along with the
groundbreaking discoveries of the Event Horizon Telescope (EHT)
\cite{sd3,sd4,sd6,sd10,jp20,jp21}, has provided the strong evidence
for the existence of BHs and their strong-field relativistic
effects, marking a pivotal moment in physical and astronomical
research. The astronomical structure of accretion disk around BHs
provided by the EHT, shows that a relatively dark central region and
a brighter outer ring structure. These two regions are known as the
shadow of the BH and the bright photon ring, respectively. This
earthshaking achievement, not only confirmed the predictions of GR
but also opened up new window for testing and refining our
understanding of gravity as well as cosmology. Since the visual
appearance and observational data of BH released by EHT, its various
observable consequences have been discussed. An astrophysical BH
provides a stable space-time geometry, yet its visual signatures can
vary in shape and radiate a spectrum of colors due to the presence
of luminous accreting material from external sources. The study of
BH shadows has a rich history dating back to the early days of GR.
In 1960, the theoretical background for describing the Schwarzschild
BH shadow was first investigated in \cite{sd11}, where a formula was
provided to calculate its angular radius. Subsequently, Bardeen
discussed the shadow of a Kerr BH and, by exploiting the
separability of the null geodesic equations, developed a mechanism
to evaluate the shadow's boundary \cite{sd12}. Despite its
importance, these achievements initially regarded the BH shadow as a
phenomenon unlikely to be observed experimentally. However, in
\cite{sd13}, the possibility of observing the BH shadow at the
center of our Milky Way galaxy was proposed, along with the
necessary experimental constraints.

Although these historical developments offer realistic descriptions
of BH shadow dynamics, there remains a strong need to further
explore their geometric properties and address the deeper
phenomenological implications through effective theoretical and
observational approaches. In 1979, Luminet proposed a computer
program to predict the gravitational field around a BH would bend
light, causing to produce the BH's shadow \cite{gkref21}. The shadow
images for Kerr space-time with thin accretion disk have been
investigated in \cite{gkref22,gkref23}. The authors in \cite{rtb23}
discussed shadows of Schwarzschild BH surrounded by a Bach-Weyl ring
with the help of the backward ray-tracing procedure. Zhong et al.
investigated the shadow of a Kerr BH in the presence of a uniform
magnetic field \cite{en27}. Using well-known theoretical models, an
extensive investigations were carried out on the shadows of BHs with
different accretion flow matters \cite{gkref26,gkref27,sd37,sd38}
and so on. Consequently, significant progress have been made in the
study of BH shadows with the help of wave optics framework
\cite{sd21,sd22,sd23,israr1}. The development of BH shadow images
marks a pivotal milestone in gravitational physics, carrying
profound scientific implications. These images offer valuable
insights into accretion processes, radiation phenomenon, and jet
formations in the vicinity of BHs, while also shedding light on the
underlying space-time structure.

Among the various frameworks and proposals to extend the study of BH
shadows through different mechanisms, the celestial light sphere
model offers a useful approach for investigating BH shadows and
light distortion around them \cite{sd24}. Hou et al. \cite{en28}
discussed the impact of rotation, electromagnetic field, and
observer inclination on the images of a rotating Kerr-Melvin BH,
where the complex structure of accretion disk model is analysed in
more depth. In the background of strong magnetic field, the authors
in \cite{en29} discussed the impact of involved parameters on the
shadow images of Kerr BH. Subsequently, the visual signatures of
various rotating BHs models illuminated by thin accretion disks were
discussed in \cite{en28,sd34,sd35,sd36,en30,en31,en32} through
effective implementations making important developments to
astronomical observations. Of course, in recent years, through the
energetic achievements of scientific community, substantial
developments have been made in the theoretical study of BHs hotspot
and polarized images \cite{en33,en34,en35,en36,jp10,jp13}, jet flow
around BHs \cite{en37,new1}, boson star's images
\cite{sd40,sd41,sd42,en38,en39} and so on. In this regard, this
paper will focus on the astronomical visual signatures of rotating
BHs in Horndeski gravity, and investigate the influence of hair
parameter on both the significant properties of disk and its image.

The segments of the present paper is as follows: In Sec. {\bf II},
we will briefly explain about the rotating BHs in Horndeski gravity
and discuss the horizon regularity, shadow contours and the shadow
observable $R_d$ and $\delta_d$, which reveals the impact of both
$a$ and $h$ on the visual characteristics of the considering BH
space-time. In Sec. {\bf III}, we investigate the visual
characteristics of the considering BH model under celestial light
source illumination. We will investigate the influence of variations
of parameters on shadow images, red-shift factors, lensing bands
with the prograde and retrograde accretion flows under thin
accretion disk model in Sec. {\bf IV}. In Sec. {\bf V}, we discuss
the constraint on the relevant parameter by the recent observational
data from M$87^{\ast}$ and Sgr $A^{\ast}$. The last section is
devoted to summarizing our conclusions.

\section{Rotating Black Holes in Horndeski Gravity}
The rotating BHs in Horndeski gravity which in the Boyer-Lindqist
coordinates is given as follows \cite{en26,en40}
\begin{eqnarray}
ds^2&=&-\bigg(\frac{\Delta-a^2\sin^2\theta}{\Sigma}\bigg)dt^2+\frac{\Sigma}{\Delta}{dr^2}+\Sigma
d\theta^2+\frac{2a\sin^2\theta}{\Sigma}\bigg(\Delta-(r^2+a^2)\bigg)dt d\varphi\nonumber\\
&+&\frac{\sin^2\theta}{\Sigma}\bigg[(r^2+a^2)^2-\Delta
a^2\sin^2\theta\bigg]d\varphi^2, \label{hd1}
\end{eqnarray}
with
\begin{equation}
\Delta=r^2+a^2-2Mr+hr\ln\big(\frac{r}{2M}\big), \quad
\Sigma=r^2+a^2\cos^2\theta, \label{hd2}
\end{equation}
where $h$ and $a$ represents the hair and spin parameters,
respectively. In the absence of hair parameter such as,
$h\rightarrow 0$, the above metric reduces to the Kerr metric and
when $a=0$, it transform to the line element of a static hairy
Horndeski BH \cite{en26}. And when both $h$ and $a$ approaches to
zero, the above metric corresponds to the Schwarszchild BH
\cite{en26}.

\begin{figure}[H]
\centering
\subfigure[\tiny][~$h=0.001$]{\label{a1}\includegraphics[width=5.4cm,height=5.2cm]{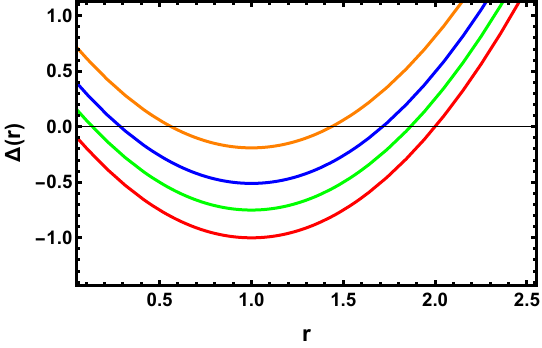}}
\subfigure[\tiny][~$h=0.5$]{\label{b1}\includegraphics[width=5.4cm,height=5.2cm]{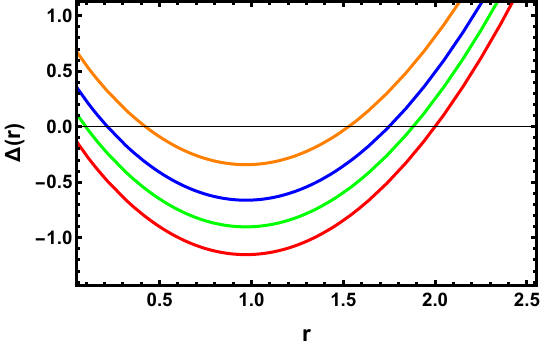}}
\subfigure[\tiny][~$h=0.9$]{\label{c1}\includegraphics[width=5.4cm,height=5.2cm]{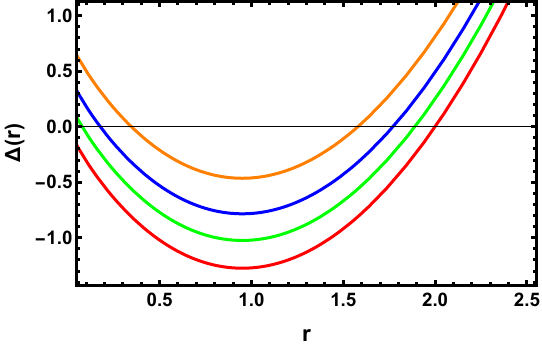}}
\caption{Physical behaviour of $\Delta(r)$ versus $r$. In all
panels, the curves red, green, blue, and orange corresponds to $a =
0.001$, $0.5$, $0.7$, and $0.9$, respectively.}\label{prd1}
\end{figure}
In Fig. \textbf{\ref{prd1}}, we analyse the horizon regularity for
$r>0$ with different values of hair parameter $h$ and spin parameter
$a$. All these results shows that there are two horizons such as the
inner horizon $r_{-}$ and the event horizon $r_{+}$. In each
trajectory, the locations where $\Delta(r)$ varies the sign
correspond to the positions of these horizons. From these results
one can notice that both horizon radii depend on the both $h$ and
$a$, as horizon radii decreases with the increasing of $a$.
Moreover, changes in $h$ slightly increase the separation between
the curves, leads to modifying the spacing between the inner and the
event horizons. The BH shadow is formed by light particles
originating from infinity, propagating toward and observed by an
arbitrary distant observer. Light rays comes from infinity are bend
towards the BH due to strong gravitational lensing and mainly
participate to the background illumination of the shadow image. In
contrast, photons that originate near the photon sphere play a
crucial role in shaping the structure of the BH shadow.
Theoretically, photons trajectories are intrinsically unstable in
the vicinity of the photon sphere and can interpret two types of
motion under perturbations such as: (i) photons that are fail to
reach the distant observer, falls into the BH produce the dark
region, so-called the BH shadow, (ii) photons that escape from the
vicinity of the photon sphere and reach distant observers outline
the outer region of the shadow, which is well-known as the critical
curve in the BH shadow image. In this way, the essential properties
of the BH space-time can be investigated through the fundamental
characteristics of the BH shadow configuration. In this scenario,
one can obtain the geodesic equations of motion from the
Hamilton-Jacobi equation formalism as given by \cite{sd60}

\begin{eqnarray}\label{hd3}
\frac{\partial \mathcal{S}}{\partial
\varpi}=-\frac{1}{2}g^{\varrho\sigma}\frac{\partial
\mathcal{S}}{\partial x^{\varrho}}\frac{\partial
\mathcal{S}}{\partial x^{\sigma}},
\end{eqnarray}
where $\mathcal{S}$ indicates the Jacobi action and $\varpi$ is the
affine parameter of the trajectory curves. Using Carter's
separability prescription \cite{sd60}, the action can be separated
into the following form
\begin{equation}\label{s4}
\mathcal{S}=\frac{1}{2}\mu^2\varpi-E
t+L\varphi+S_r(r)+S_{\theta}(\theta),
\end{equation}
in which $\mu=0$ stands for the rest mass of the particles. The
constants $E=-p_{t}$ and $L=p_{\varphi}$ interpret the conserved
energy and conserved angular momentum of the photon in the direction
of rotation axis, respectively. The functions $S_r(r)$ and
$S_{\theta}(\theta)$ are arbitrary functions.

\begin{figure}[H]\centering
\subfigure[\tiny][]{\label{alb1}\includegraphics[width=8cm,height=7cm]{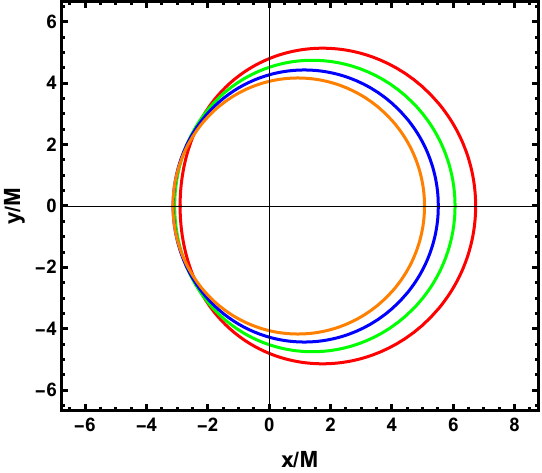}}
\caption{The shadow contours are plotted for fixed values of $a =
0.9$ and $\theta_{\text{obs}}=70^{\circ}$, corresponding to
different values of the parameter $h$. The red, green, blue, and
orange circular orbits corresponds to $h=0.1$,~$1.1$,~$2.1$, and
$3.1$, respectively.}\label{prd2}
\end{figure}

Upon implementing into account the Horndeski BH space-time, the
equations of motion according to the four differential equations can
be defined as
\begin{eqnarray}\nonumber
\Sigma^{2}\frac{dt}{d\varpi}&=&a(L-aE\sin^{2}\theta)+\frac{r^{2}+a^{2}}{\Delta}(E(r^{2}+a^{2})-a
L),\\\nonumber
\Sigma^{2}\frac{dr}{d\varpi}&=&\pm\sqrt{\hat{R}(r)},\\\nonumber
\Sigma^{2}\frac{d\theta}{d\varpi}&=&\pm\sqrt{\Theta(\theta)},\\\label{s5}
\Sigma^{2}\frac{d\phi}{d\varpi}&=&(L\csc^{2}\theta-aE)+\frac{a}{\Delta}(E(r^{2}+a^{2})-aL),
\end{eqnarray}
with
\begin{eqnarray}\nonumber
\hat{R}(r)&=&(E(r^{2}+a^{2})-aL)^{2}-\Delta(Q+(L-aE)^{2}),\\\label{s6}
\Theta(\theta)&=&Q+\big(a^{2}E^{2}-L^{2}\csc^{2}\theta\big)\cos^{2}\theta,
\end{eqnarray}
where $Q$ is the Carter constant. At the position of the photon
sphere, photon should satisfy $\dot{r}=0=\ddot{r}$ (the sign ``dot''
represent the derivative with respect to $\varpi$), which is
equivalent to $\hat{R}(r)=0=\partial_{r}\hat{R}(r)$ and
$\partial^{2}_{r}\hat{R}(r)\leq0$. Now the impact parameters $E$ and
$L$ has the following relations as

\begin{eqnarray}\label{s6}
\xi=\frac{L}{E}, \quad \quad \eta = \frac{Q}{E^2}.
\end{eqnarray}
In this regard, one can the obtain the values of critical impact
parameters as
\begin{eqnarray}\label{s7}
\xi(r)&=&\frac{(a^{2}+r^{2})\Delta'(r)-4r\Delta(r)}{a\Delta'(r)}\mid_{r=r_{p}},\\\label{s8}
\eta(r)&=&\frac{r^{2}(-16\Delta(r)^{2}-r^{2}\Delta'(r)^{2}+8\Delta(r)(2a^{2}+
r\Delta'(r)))}{a^{2}\Delta'(r)^{2}}\mid_{r=r_{p}},
\end{eqnarray}
within the photon region and ($'$) is the derivative with respect to
$r$. The radial intensity of the photon region beyond the event
horizon of considering BH model is calculated by the roots of the
equation $\eta(r)=0$. Solving this equation gives two largest
positive roots, which is prograde radius $r_{p}^{-}$ (where $r_{p}$
is the radius of photon sphere) and retrograde radius $r_{p}^{+}$,
indicates the inner and outer boundaries of the radial range of
unstable circular photon orbits, respectively. The photon region is
further characterized by the constraint $\Theta(\theta)\geq0$ for
spherical photon orbits. For an observer, which is lies at infinity,
can be defined as a zero-angular-momentum observer (ZAMO) at
coordinates ($t_{obs}=0,~r_{obs},~\theta_{obs},~\varphi_{obs}=0$) by
assuming the symmetries in the directions of $t$ and $\varphi$.
Hence, one can observe the optical appearance of BH shadow on the
observer screen with the help of fisheye lens camera model. In this
regard, we closely followed the outlined as defined in
\cite{sd35,sd62,new2}, where the schematic representation of the
ZAMO coordinates is provided and the procedure of the stereographic
projection technique is being used. Moreover, the schematic
representation of the discretized image plane along with the field
of view of camera, which is essential for capturing the BH shadow
image on the screen are defined in detail. For a comprehensive
review, the readers can see the Refs. \cite{sd35,sd62,new2}. The
relationship between photon $4$-momentum and the celestial
coordinates $(\alpha,~\beta)$ are expressed as \cite{sd35,sd62,new2}
\begin{eqnarray}\label{s9}
\cos\alpha=\frac{p^{(1)}}{p^{(0)}},\quad
\tan\beta=\frac{p^{(3)}}{p^{(2)}}.
\end{eqnarray}
The observer's screen can be equipped with a standard Cartesian
coordinate system $(x,~y)$, which is accurately aligned with the
celestial coordinates as
\begin{eqnarray}\label{s10}
x(r_{p})=-2\tan\frac{\alpha}{2}\sin\beta, \quad
y(r_{p})=-2\tan\frac{\alpha}{2}\cos\beta.
\end{eqnarray}
In Fig. \textbf{\ref{prd2}}, we have interpreted the circular orbits
of BH shadow for different values of hair parameter $h$. Here, we
observe that as the values of $h$ decrease, the circular orbits of
the BH shadow shift significantly towards the right side of the
screen. Moreover, with decreasing $h$, the spacing between the
circular curves becomes more noticeable on the right side. On the
other hand, on the left side of the screen, all the curves overlap,
and the circular shape of the shadow contour becomes slightly
deformed into a D-shape when $h=0.1$. Thus, the larger values of $h$
results in a substantial decrease in the shadow size.

In order to define the size and distortion of a BH shadow, one can
define the two observable quantities, which is known as shadow
radius $R_{d}$ and the deviation from circularity $\delta_{d}$.
Here, we adopt the method as proposed in \cite{sd26}, which is
defined as

\begin{eqnarray}\label{ned1}
R_{d}=\frac{(x_{t}-x_{r})^{2}+y_{t}^{2}}{2\mid x_{t}-x_{r}\mid},
\quad \delta_{d}=\frac{\mid x_{l'}-x_{l}\mid}{R_{d}}.
\end{eqnarray}

The top, bottom, and rightmost positions of the BH shadow
individually determine a reference circle, whose radius is $R_{d}$,
approximating the size of the shadow. The parameter $\delta_{d}$
represents the absolute horizontal difference between the leftmost
points of the BH shadow and the reference circle, indicating the
degree of deviation from circularity. Particularly, the five
reference points ($x_{t}, y_{t}$), ($x_{b}, y_{b}$), ($x_{r}, 0$),
($x_{l}, 0$), and ($x_{l'}, 0$) correspond to the top, bottom,
rightmost, and leftmost points of the shadow, leftmost point of the
reference circle, respectively. When $x_{l}\neq x_{l'}$,
$\delta_{d}\neq0$. The larger value of $\delta_{d}$ represents a
greater deviation of the BH shadow boundary from a circular size.

\begin{figure}[H]
\begin{center}
\subfigure[\tiny][$~a=0.5$]{\label{a1}\includegraphics[width=7cm,height=7cm]{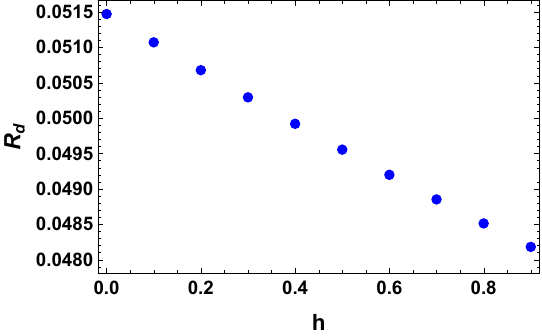}}
\subfigure[\tiny][$~a=0.5$]{\label{b1}\includegraphics[width=7cm,height=7cm]{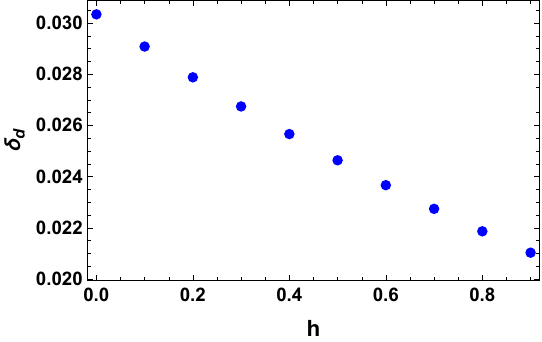}}
\subfigure[\tiny][$~h=0.5$]{\label{c1}\includegraphics[width=7cm,height=7cm]{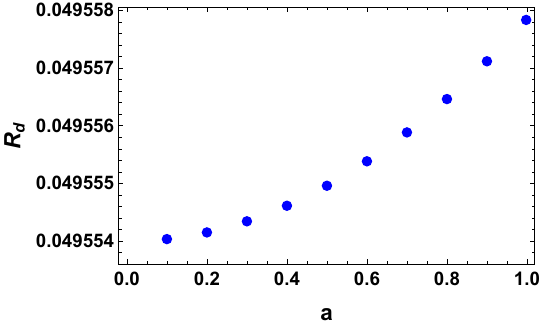}}
\subfigure[\tiny][$~h=0.5$]{\label{d1}\includegraphics[width=7cm,height=7cm]{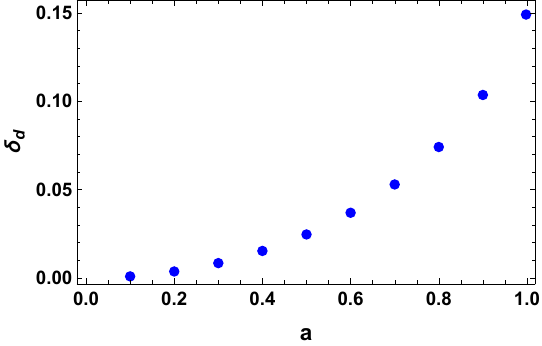}}
\caption{The shadow observable $R_d$ and $\delta_d$ for
$r_{obs}=100$, $\theta_{obs}=70^{\circ}$ and $M=1$.}\label{prd3}
\end{center}
\end{figure}
From the top row of Fig. \textbf{\ref{prd3}}, we observe that the
shadow radius $R_d$ and deviation $\delta_d$ are decreases smoothly
with the increasing values of $h$. These results are also consistent
with Fig. \textbf{\ref{prd2}}, where one can observe that the larger
values of $h$, results in decrease the shadow radius as well as the
corresponding distortion. We also plot the shadow observable $R_d$
and $\delta_d$ with the variations of $a$, see bottom row of Fig.
\textbf{\ref{prd3}}. Here one can observe that both $R_d$ and
$\delta_d$ are increases with the increasing values of $a$.
Moreover, here is a little influence of $a$ on both quantities.

\section{Celestial Light Source Illumination}
\begin{figure}[H]
\begin{center}
\subfigure[\tiny][~$a=0.001,~h=0.001$]{\label{a1}\includegraphics[width=3.9cm,height=4cm]{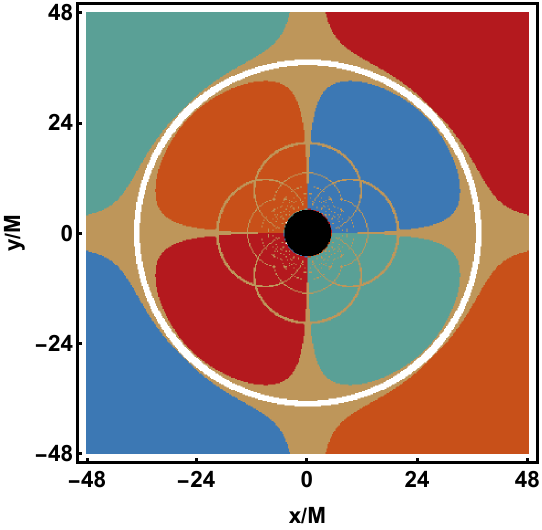}}
\subfigure[\tiny][~$a=0.001,~h=0.1$]{\label{b1}\includegraphics[width=3.9cm,height=4cm]{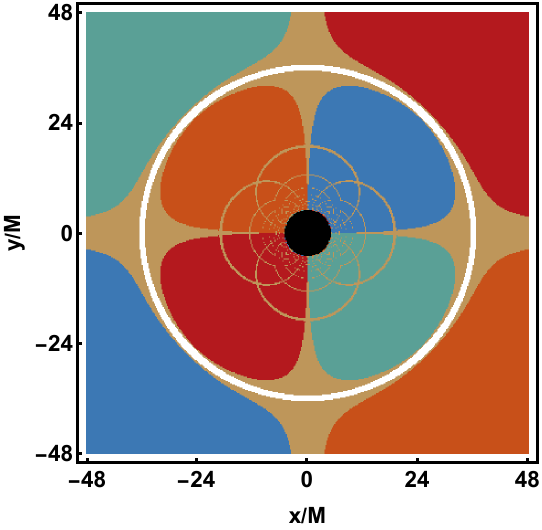}}
\subfigure[\tiny][~$a=0.001,~h=0.5$]{\label{c1}\includegraphics[width=3.9cm,height=4cm]{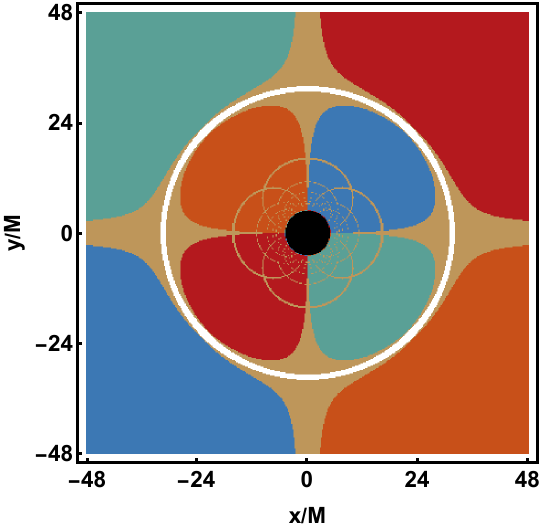}}
\subfigure[\tiny][~$a=0.001,~h=0.9$]{\label{d1}\includegraphics[width=3.9cm,height=4cm]{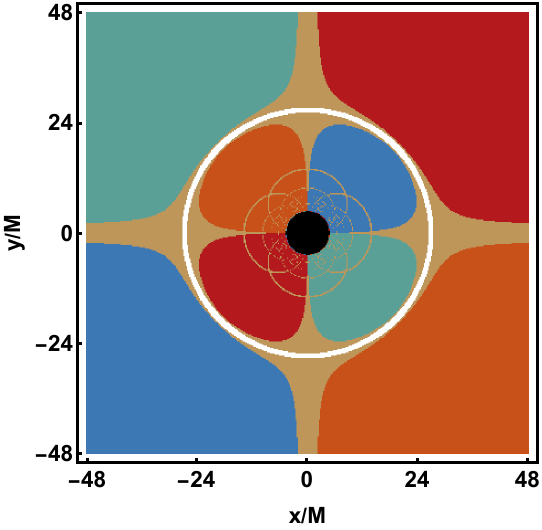}}
\subfigure[\tiny][~$a=0.5,~h=0.001$]{\label{a2}\includegraphics[width=3.9cm,height=4cm]{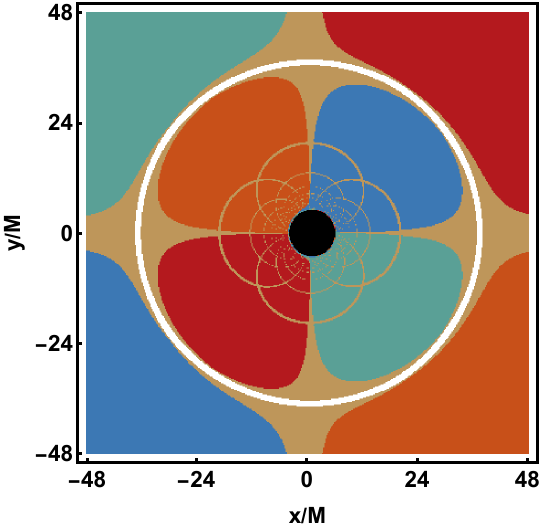}}
\subfigure[\tiny][~$a=0.5,~h=0.1$]{\label{b2}\includegraphics[width=3.9cm,height=4cm]{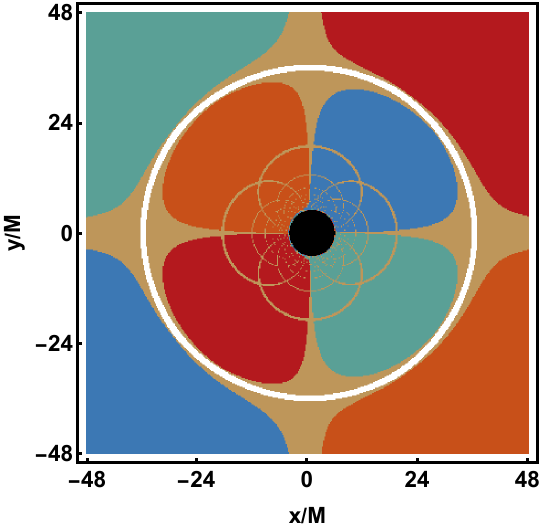}}
\subfigure[\tiny][~$a=0.5,~h=0.5$]{\label{d2}\includegraphics[width=3.9cm,height=4cm]{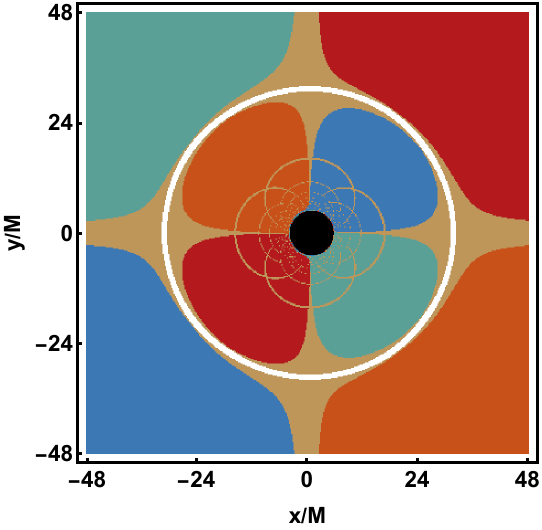}}
\subfigure[\tiny][~$a=0.5,~h=0.9$]{\label{d2}\includegraphics[width=3.9cm,height=4cm]{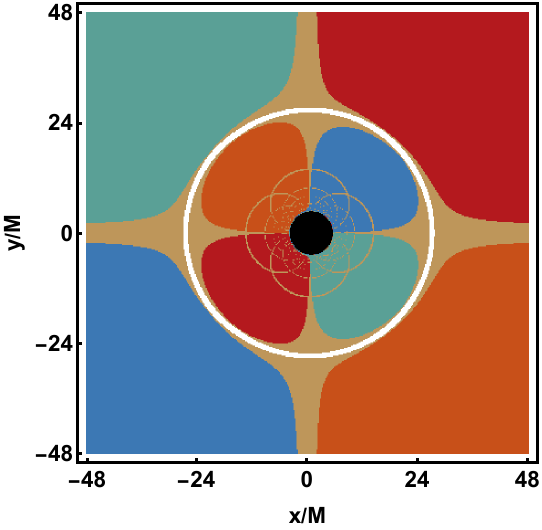}}
\subfigure[\tiny][~$a=0.9,~h=0.001$]{\label{a3}\includegraphics[width=3.9cm,height=4cm]{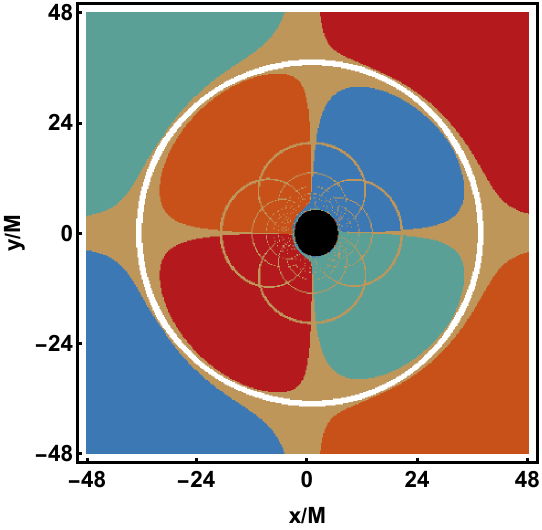}}
\subfigure[\tiny][~$a=0.9,~h=0.1$]{\label{b3}\includegraphics[width=3.9cm,height=4cm]{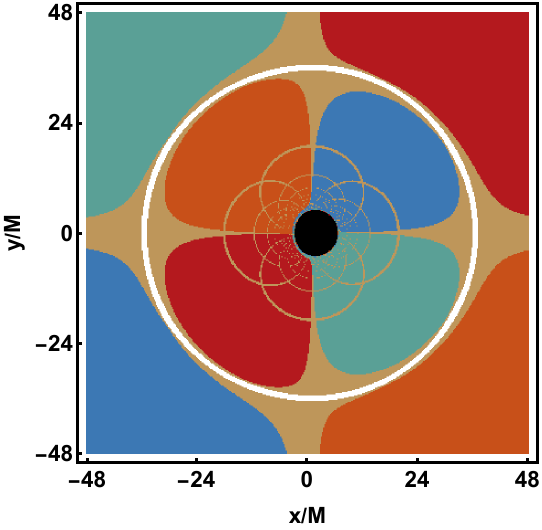}}
\subfigure[\tiny][~$a=0.9,~h=0.5$]{\label{c3}\includegraphics[width=3.9cm,height=4cm]{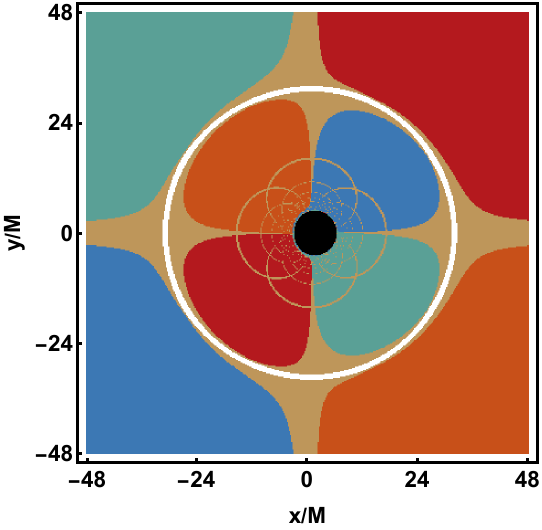}}
\subfigure[\tiny][~$a=0.9,~h=0.9$]{\label{d3}\includegraphics[width=3.9cm,height=4cm]{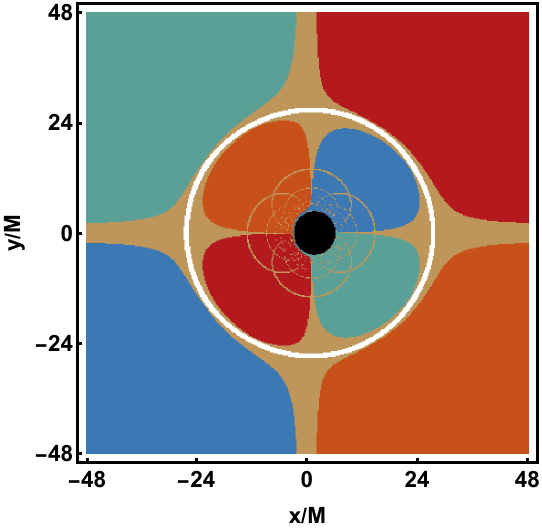}}
\caption{Optical images of rotating BHs in Horndeski gravity with
different values of $a$ and $h$ with fixed
$\theta_{obs}=80^{\circ}$. The grids of the longitude and latitude
lines are marked with adjacent brown lines separated by
$10^{\circ}$.}\label{prd4}
\end{center}
\end{figure}

Now, we consider the backward ray-tracing technique to investigate
the optical characteristics of the rotating BHs in Horndeski gravity
with a celestial light source model. In this scenario, the central
dark region shows the BH shadow, with its rotation axis pointing
toward the north pole. The size of the BH is much smaller as compare
to the diameter of the celestial sphere. For better physical
understanding, one can equally divide the celestial sphere into four
different quadrants, assigned with red, orange, blue, and cyan,
respectively. The spherical longitude and latitude lines is
indicated by brown marker positioned by $10^{\circ}$. In each panel,
outside the ``D'' shape petals, there is a white circular ring which
could provide a direct interpretation of Einstein ring. Obviously,
all these images clearly reflects the warping of space by a BH and
gravitational lensing influence of a BH. Closely followed by the
strategy as defined in \cite{sd62}, we consider the fish eye camera
model and obtain the BH shadow images for different values of $a$
and $h$, with fixed inclination angle $\theta_{obs}=80^{\circ}$, see
Fig. \textbf{\ref{prd4}}.

From the upper row of Fig. \textbf{\ref{prd4}}, it can be noticed
that when both $a\rightarrow0$ and $h\rightarrow0$, the BH shadow
appears as a perfect circle and the intersection line between the
orange and blue colours inside the Einstein ring is perpendicular to
the x-axis, indicating that the BH is approximately static, with no
significant space-dragging effect. As the values of $h$ are
increases, the ``D'' shape petals are slightly evolves and the
radius of the white circular ring are gradually moves towards the
interior of the BH. However, the size of the solid black disk
remains almost same in all cases. In the second case, when we
increase the value of $a$ such as $a=0.5$ and varies $h$ from left
to right, we notice that the intersection line between the orange
and blue colours inside the Einstein ring is not exactly
perpendicular to the x-axis, which represents the significant
influence of space-dragging effect. Moreover, with the increasing of
$h$, one can observe that the size of the shadow contour as well as
the radius of the Einstein ring are decreases significantly.
Similarly, when we further increase the value of $a$, the
space-dragging effect are more obvious as compare to previous two
cases, see third row of Fig. \textbf{\ref{prd4}}. In this case, one
can also observe the the shadow contours as well as the radius of
the Einstein ring are decreases significantly with increasing values
of $h$. Therefore, the impact of $a$ leads to increase the
space-dragging effect, while the increasing values of $h$, results
in decrease the shadow contours as well as the corresponding
Einstein ring radius.

\section{Thin Accretion Disk Illumination}
It is well-known that, the light source of a BH is usually the
accretion disk surrounding it. Therefore, investigating the visual
characteristics of rotating BHs under the accretion disk
illumination provides the significant information about the
intricate properties of BH shadows. In this view, one can assume
that the accretion disk surrounding the rotating BH in Horndeski
gravity is optically and geometrically thin, lies at the equatorial
plane. And the observer position lies at a sufficiently large
distance. The accretion disk consists of freely moving, electrically
neutral plasma following time-like geodesics along the equatorial
plane. For better understanding about the phenomenon of accreting
matter close to the BH, we divide the accretion flow of accreting
matter into two portions such as the innermost stable circular orbit
(ISCO), where the matter falls into the event horizon along plunging
orbits, and the region beyond the ISCO, where matter moves along
Keplerian orbits.

In this study, we assume that the accretion disk starts from the BH
event horizon $r_{+}$ and extends outward to a sufficiently large
distance, name as $r_{f}$. The observer's radial position satisfies
the region $r_{+}\ll r_{obs}<r_{f}$. In order to discuss the visual
characteristics of the considering BH model, we first determine the
position of ISCO, with the help of the following equations
\cite{en28}

\begin{eqnarray}\label{s11}
V_{eff}(r)=0, \quad \partial_{r} V_{eff}(r)=0, \quad
\partial_{r}^{2} V_{eff}(r)=0,
\end{eqnarray}
where $V_{eff}$ is the effective potential, which define as
\begin{equation}\label{s12}
V_{eff}=(1+g^{tt}\hat{E}^{2}+g^{\varphi\varphi}\hat{L}^{2}-2g^{t\varphi}\hat{E}\hat{L}),
\end{equation}
with
\begin{eqnarray}\label{s13}
\hat{E}=-\frac{1}{\sqrt{f_{1}}}(g_{tt}+g_{t\varphi}\widetilde{\omega}),\quad
\hat{L}=\frac{1}{\sqrt{f_{1}}}(g_{t\varphi}+g_{\varphi\varphi}\widetilde{\omega}),
\end{eqnarray}
where
$f_{1}=-g_{tt}-2g_{t\varphi}\widetilde{\omega}-g_{\varphi\varphi}\widetilde{\omega}^{2}$
and
$\widetilde{\omega}=\frac{d\varphi}{dt}=(\partial_{r}g_{t\varphi}+(\sqrt{\partial^{2}_{r}g_{t\varphi}-\partial_{r}g_{tt}\partial_{r}g_{\varphi}})(\partial_{r}g_{\varphi\varphi})^{-1}.$
When $r=r_{ISCO}$, there are two conserved quantities, represented
as $\hat{E}_{ISCO}$ and $\hat{L}_{ISCO}$. When $r>r_{ISCO}$, the
matter in the accretion disk moves along Keplerian orbits, with its
four-velocity given by
\begin{equation}\label{s14}
K^{\xi}_{out}=\frac{1}{\sqrt{f_{1}}}(1,0,0,\widetilde{\omega}).
\end{equation}
On the other hand, within the ISCO, the accretion flows descend from
the ISCO to the event on a critical plunging orbits, preserving the
conserved quantities related to the ISCO. In this scenario, the
components of four-velocity are defined as \cite{en28}
\begin{eqnarray}\nonumber
K^{t}_{plung}&=&(-g^{tt}\hat{E}_{ISCO}+g^{t\varphi}\hat{L}_{ISCO}),\quad
K^{\varphi}_{plung}=(-g^{t\varphi}\hat{E}_{ISCO}+g^{\phi\phi}\hat{L}_{ISCO}),\\\nonumber
K^{r}_{plung}&=&-\big(-(g_{tt}K^{t}_{plung}K^{t}_{plung}+2g_{t\varphi}K^{t}_{plung}
K^{\varphi}_{plung}+g_{\varphi\varphi}K^{\varphi}_{plung}K^{\varphi}_{plung}+1)(g_{rr})^{-1}\big)^{\frac{1}{2}},\\\label{s15}~K^{\theta}_{plung}&=&0.
\end{eqnarray}
Since, the light particles may crossed the accretion disk plane once
($n=1$), twice ($n=2$) or even many times ($n>2$), which is
corresponds to the direct, lensed or higher-order images,
respectively. In this study, we focus exclusively on two cases:
direct and lensed images. As it is well known that, when a light
particle intersects the accretion disk, variations in its intensity
are primarily caused by photon emission and absorption. For
simplicity, reflection effects are neglected in the present model.
Therefore, the observed intensity on the observer's screen can be
expressed as \cite{en28}
\begin{equation}\label{s16}
\mathcal{I}_{obs}=\sum_{n=1}^{N_{max}}f_{n}\Psi_{n}^{3}(r_{n})\Gamma_{n},
\end{equation}
Here $n=1,2,3...N_{max}$ indicates the number of times a light ray
crossed the equatorial plane and $f_{n}=1$ is a fudge factor. The
parameter $\Psi_{n}=\nu_{obs}/\nu_{n}$ is the red-shift factor, in
which $\nu_{obs}$ is the observed frequency by the observer,
$\nu_{n}$ is the frequency measured in the local rest frame comoving
with the accretion disk. The expression of $\Gamma_{n}$ is
second-order polynomial in logarithmic space, which is defined as
\begin{equation}\label{s17}
\Gamma_{n}=\exp\big[\rho_{1}k^{2}+\rho_{2}k\big],
\end{equation}
where $k=\log(\frac{r}{r_{+}})$ and $\rho_{1}=-1/2$ and
$\rho_{2}=-2$ \cite{en27}. Naturally, the red-shift factor exhibits
distinct functional forms in the inner and outer regions of the
ISCO, reflecting the pronounced differences in particle emission
spectra between these two domains. The red-shift factor outside the
ISCO can be expressed as \cite{en27}
\begin{equation}\label{s18}
\Psi^{out}_{n}=\frac{\tau(1-\lambda\frac{p_{\varphi}}{p_{t}})}{\sigma(1+\widetilde{\omega}\frac{p_{\varphi}}{p_{t}})}|_{r=r_{n}},
\quad \quad r\geq r_{ISCO},
\end{equation}
where
$\tau=\sqrt{\frac{g_{\varphi\varphi}}{g^2_{t\varphi}-g_{tt}g_{\varphi\varphi}}}$,~$\lambda=\frac{g_{t\varphi}}{g_{\varphi\varphi}}$,~
$\sigma=\frac{1}{\sqrt{f_{1}}}$ and
$\bar{e}=\frac{p_{(t)}}{p_{t}}=\tau(1-\lambda\frac{p_{\varphi}}{p_{t}})$
is the relationship between the observed energy on the screen to the
energy along a null geodesic. For an asymptotically flat space-time,
when the observer is at infinity, we have $\bar{e}=1$. When $r<
r_{ISCO}$, the accretion flow is moving along the critical plunging
orbit, then the red-shift factor is expressed as \cite{en27}
\begin{equation}\label{s19}
\Psi^{plung}_{n}=-\frac{1}{K^{r}_{plung}p_{r}/p_{t}-\hat{E}_{ISCO}(g^{tt}-g^{t\varphi}p_{\varphi}/p_{t})
+\hat{L}_{ISCO}(g^{\varphi\varphi}p_{\varphi}/p_{t}+g^{t\varphi})}|_{r=r_{n}},
\quad \quad r< r_{ISCO}.
\end{equation}

\begin{figure}[H]
\begin{center}
\subfigure[\tiny][~$a=0.001,~h=0.001$]{\label{a1}\includegraphics[width=3.9cm,height=3.7cm]{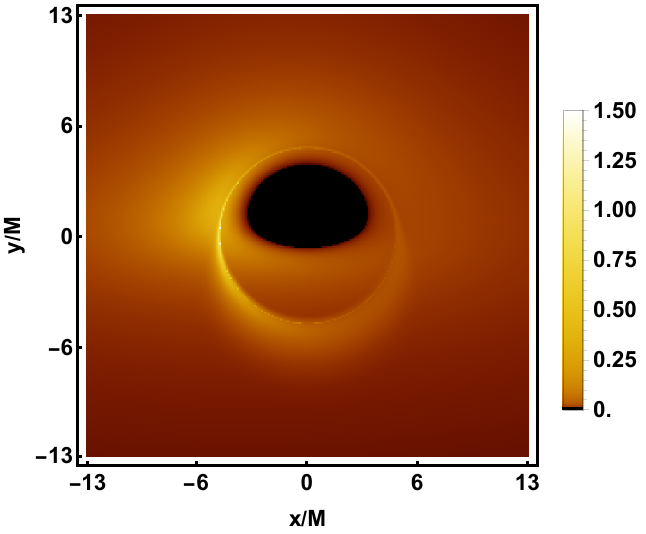}}
\subfigure[\tiny][~$a=0.001,~h=0.1$]{\label{b1}\includegraphics[width=3.9cm,height=3.7cm]{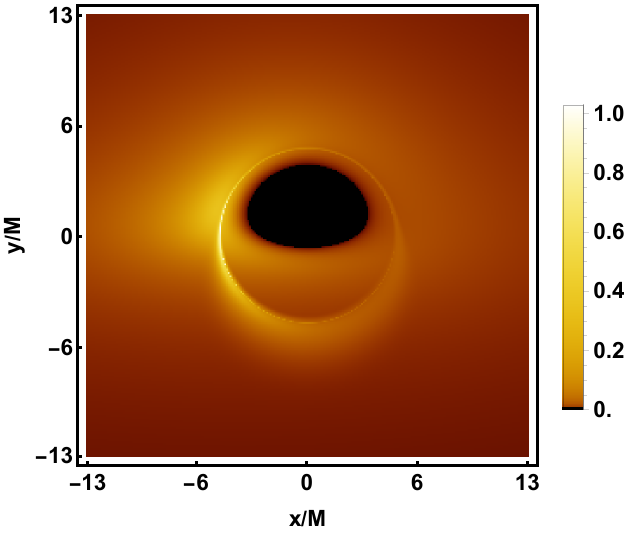}}
\subfigure[\tiny][~$a=0.001,~h=0.5$]{\label{c1}\includegraphics[width=3.9cm,height=3.7cm]{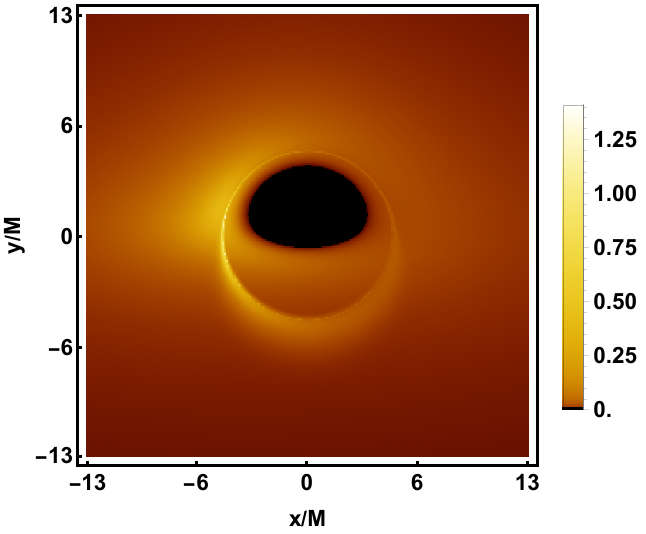}}
\subfigure[\tiny][~$a=0.001,~h=0.9$]{\label{d1}\includegraphics[width=3.9cm,height=3.7cm]{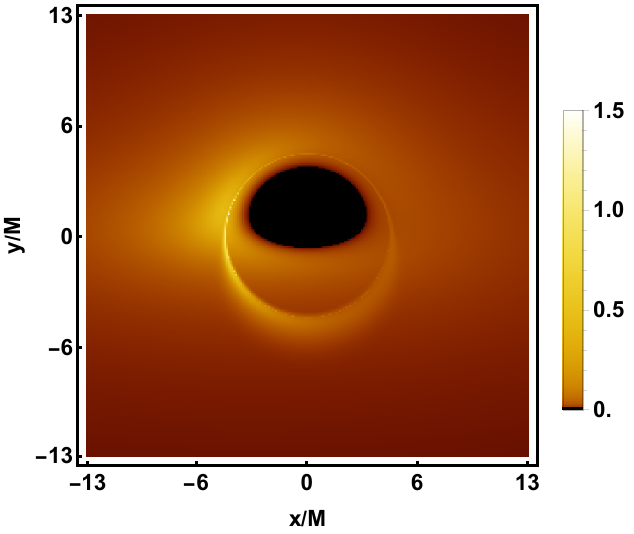}}
\subfigure[\tiny][~$a=0.5,~h=0.001$]{\label{a2}\includegraphics[width=3.9cm,height=3.7cm]{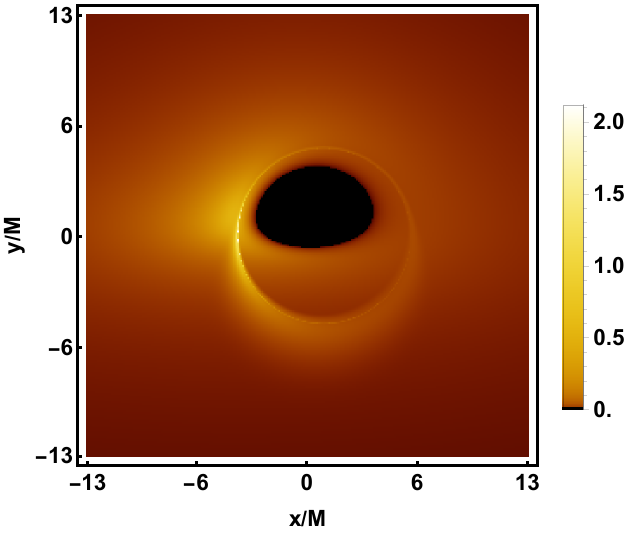}}
\subfigure[\tiny][~$a=0.5,~h=0.1$]{\label{b2}\includegraphics[width=3.9cm,height=3.7cm]{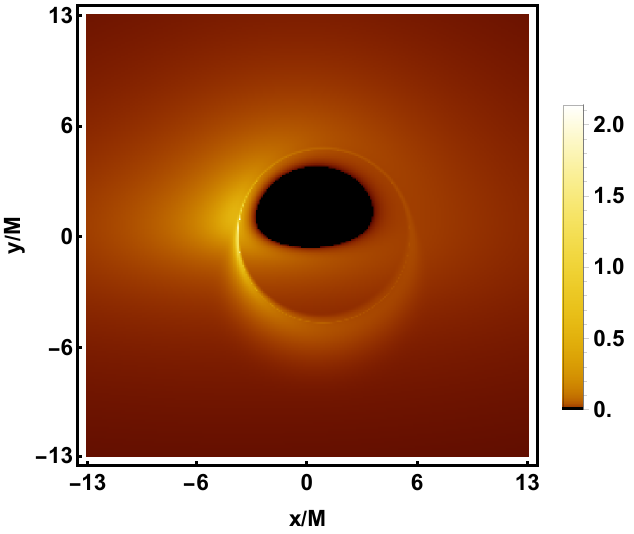}}
\subfigure[\tiny][~$a=0.5,~h=0.5$]{\label{d2}\includegraphics[width=3.9cm,height=3.7cm]{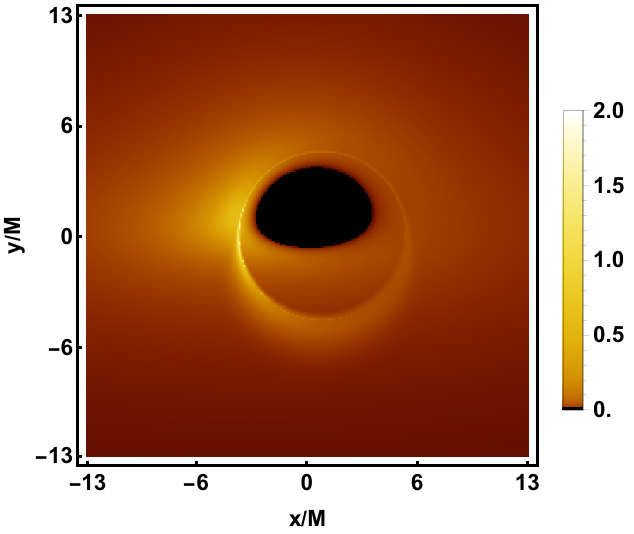}}
\subfigure[\tiny][~$a=0.5,~h=0.9$]{\label{d2}\includegraphics[width=3.9cm,height=3.7cm]{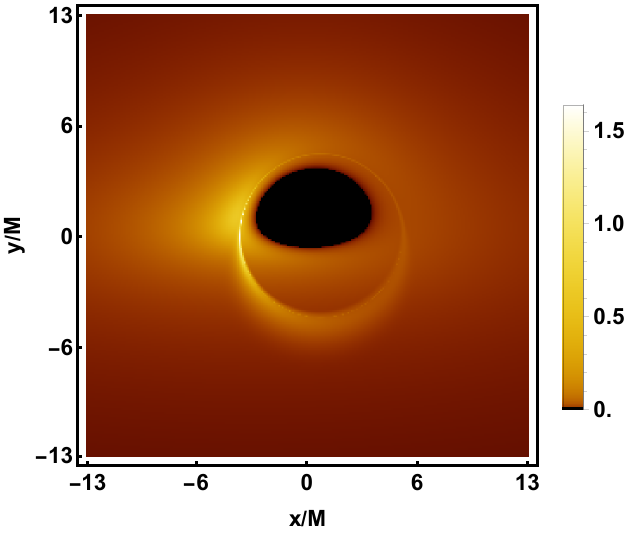}}
\subfigure[\tiny][~$a=0.9,~h=0.001$]{\label{a3}\includegraphics[width=3.9cm,height=3.7cm]{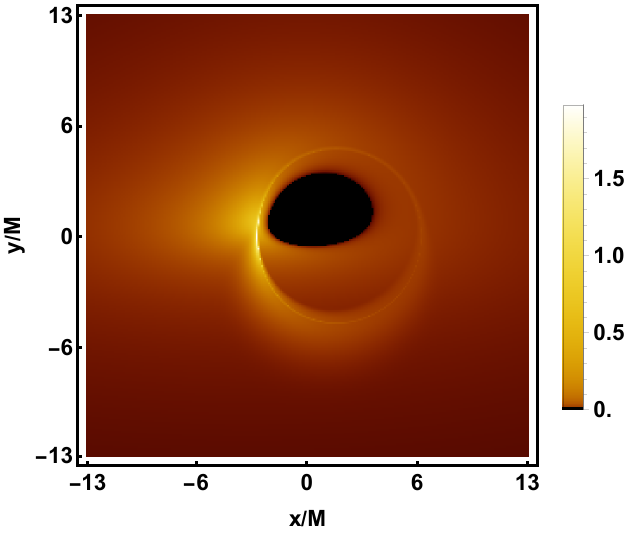}}
\subfigure[\tiny][~$a=0.9,~h=0.1$]{\label{b3}\includegraphics[width=3.9cm,height=3.7cm]{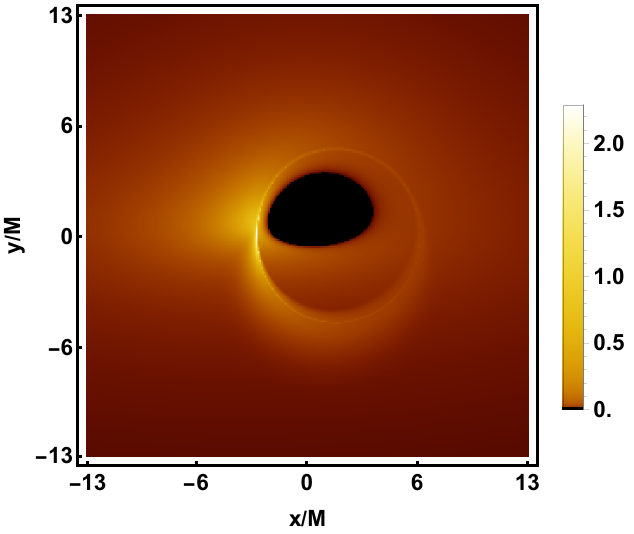}}
\subfigure[\tiny][~$a=0.9,~h=0.5$]{\label{c3}\includegraphics[width=3.9cm,height=3.7cm]{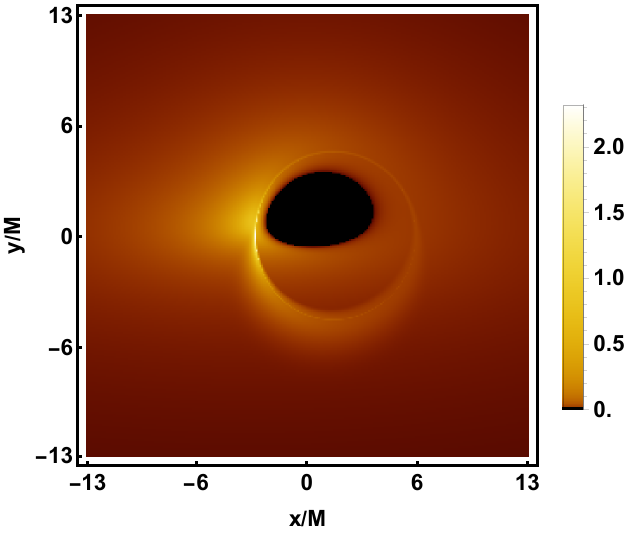}}
\subfigure[\tiny][~$a=0.9,~h=0.9$]{\label{d3}\includegraphics[width=3.9cm,height=3.7cm]{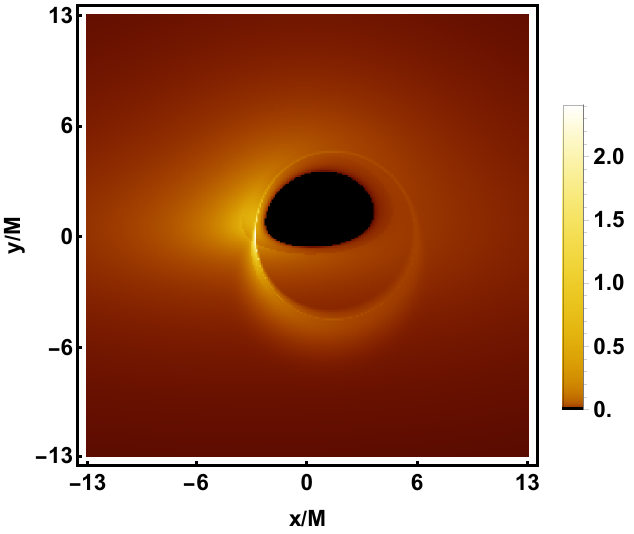}}
\caption{Optical images of rotating BHs in Horndeski gravity with
different values of $a$ and $h$ with fixed $\theta_{obs}=70^{\circ}$
under prograde flow. The BH's event horizon is represented as a
black region and a luminous circular ring corresponds to the
position of the bright photon ring.}\label{prd5}
\end{center}
\end{figure}
In Fig. \textbf{\ref{prd5}}, we depict the impact of the rotation
parameter $a$ and the hair parameter $h$ with fixed
$\theta_{obs}=80^{\circ}$ within the framework of prograde flow. It
can be noticed that, regardless of the variations in both $a$ and
$h$, the central black region always exist in the middle of the
screen, which is corresponds to the zone where photons are directly
fall into the BH. Moreover, there is always a bright closed curve at
the outer boundary of the inner shadow, which is known as the
critical curve \cite{perl} and the size of the inner shadow appears
as a ``D'' shape in each panel. In order to observe the influence of
hair parameter $h$, we fixed the rotation parameter $a$ and vary the
values of $h$ from left to right as $0.001,~0.1,~0.5,~0.9$. Whereas
from top to bottom, the intensity maps illustrates the impact of the
$a$ on the images of BH shadows. From these images, it is noticed
that with the alteration of the $h$, the shadow images are slightly
deformed, emerging a smooth, hat-like black region. Specifically,
when $a$ is fixed and $h$ is increased, it shows that the inner
shadow are slightly contracts and luminosity moves towards the upper
half-portion of the screen. Whereas, when we fixed $h$ and varies
$a$ from top to bottom, the size of the inner shadow are
significantly decreases and a ``crescent-shaped'' bright region
appears on the left side of the critical curve, significantly
increases the corresponding intensity with the larger values of $a$.
This phenomenon arises from the Doppler effect induced by the
relative motion between the prograde accretion disk and the
observer. In this configuration, light emitted from the left side of
the disk propagates toward the observer, undergoing a blue-shift
that enhances photon energy and consequently brightens the
corresponding region.

\begin{figure}[H]
\begin{center}
\subfigure[\tiny][~$a=0.001,~h=0.001$]{\label{a1}\includegraphics[width=3.9cm,height=4cm]{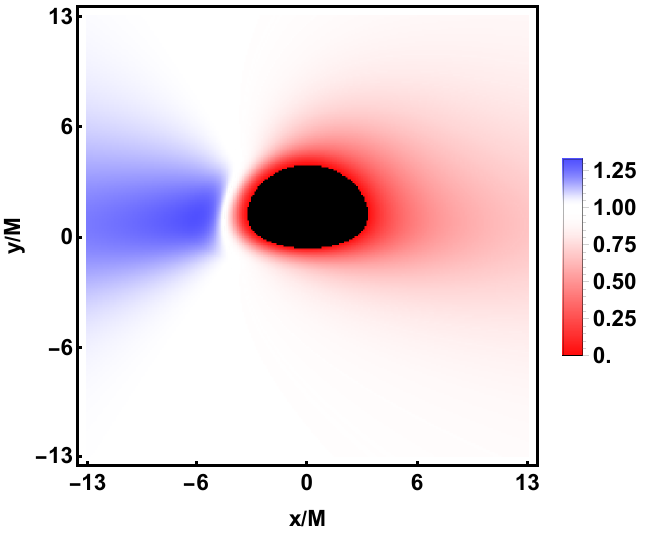}}
\subfigure[\tiny][~$a=0.001,~h=0.1$]{\label{b1}\includegraphics[width=3.9cm,height=4cm]{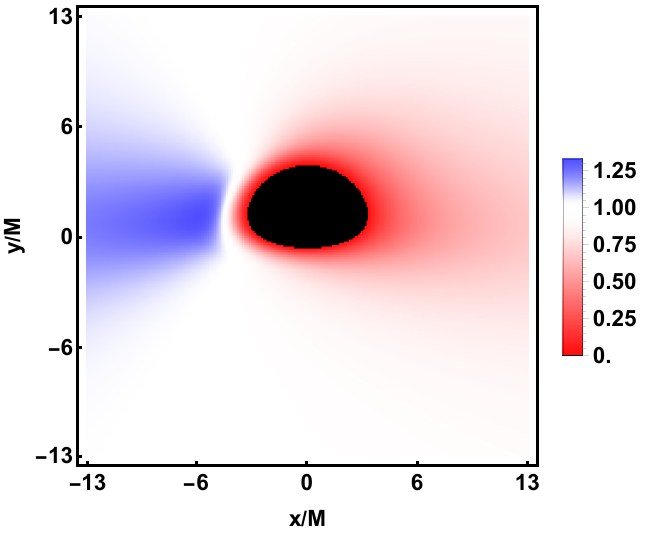}}
\subfigure[\tiny][~$a=0.001,~h=0.5$]{\label{c1}\includegraphics[width=3.9cm,height=4cm]{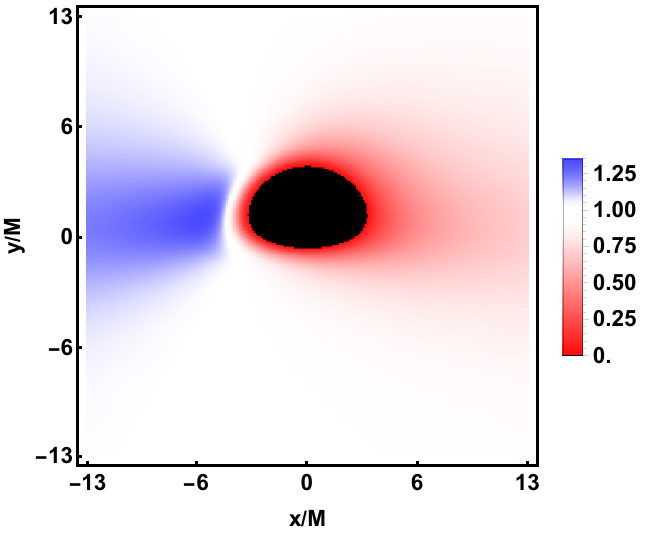}}
\subfigure[\tiny][~$a=0.001,~h=0.9$]{\label{d1}\includegraphics[width=3.9cm,height=4cm]{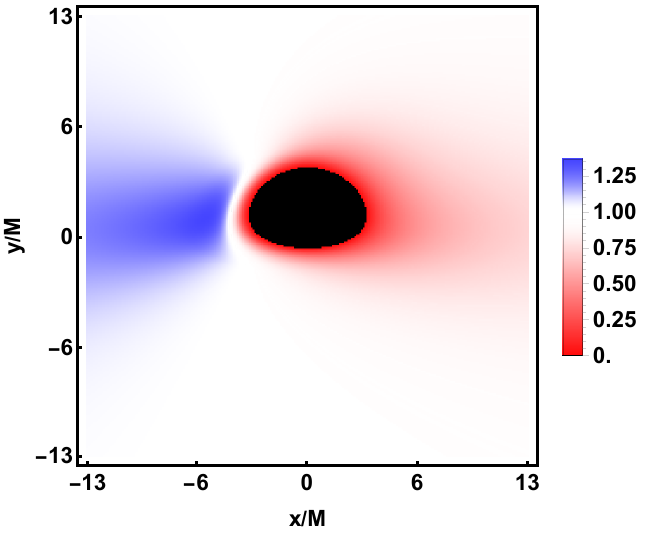}}
\subfigure[\tiny][~$a=0.5,~h=0.001$]{\label{a2}\includegraphics[width=3.9cm,height=4cm]{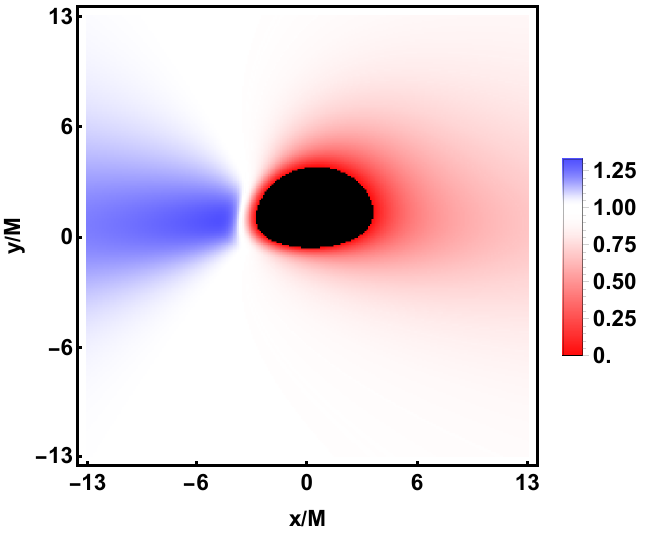}}
\subfigure[\tiny][~$a=0.5,~h=0.1$]{\label{b2}\includegraphics[width=3.9cm,height=4cm]{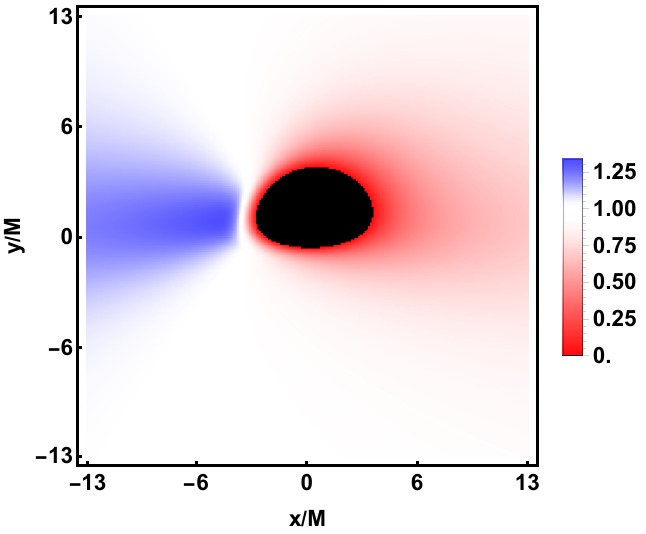}}
\subfigure[\tiny][~$a=0.5,~h=0.5$]{\label{d2}\includegraphics[width=3.9cm,height=4cm]{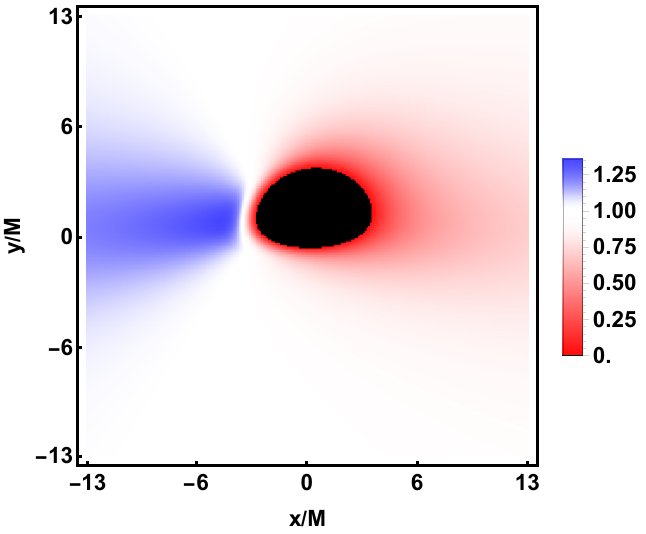}}
\subfigure[\tiny][~$a=0.5,~h=0.9$]{\label{d2}\includegraphics[width=3.9cm,height=4cm]{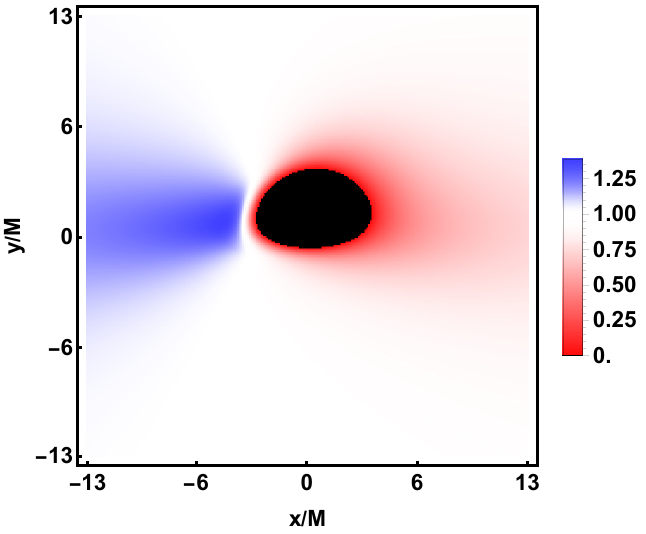}}
\subfigure[\tiny][~$a=0.9,~h=0.001$]{\label{a3}\includegraphics[width=3.9cm,height=4cm]{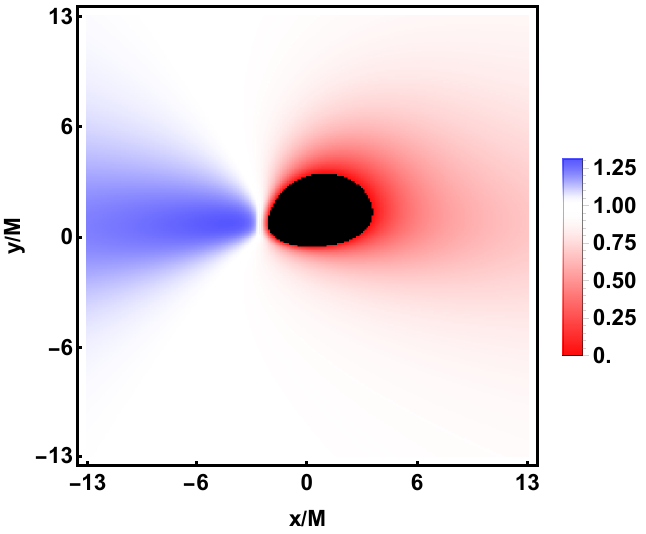}}
\subfigure[\tiny][~$a=0.9,~h=0.1$]{\label{b3}\includegraphics[width=3.9cm,height=4cm]{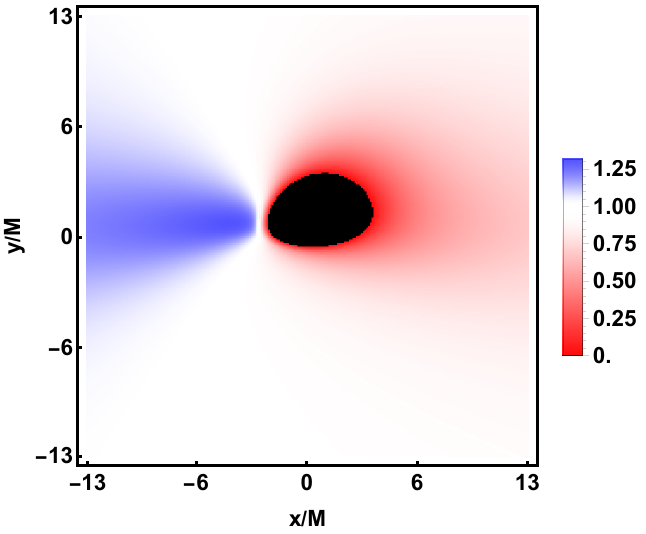}}
\subfigure[\tiny][~$a=0.9,~h=0.5$]{\label{c3}\includegraphics[width=3.9cm,height=4cm]{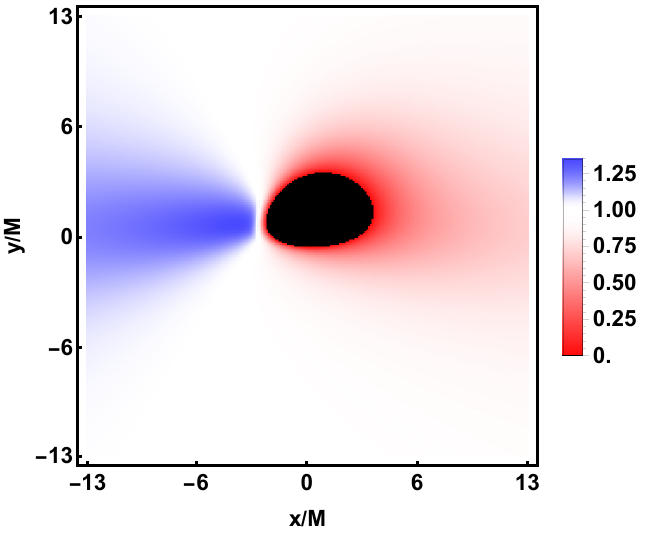}}
\subfigure[\tiny][~$a=0.9,~h=0.9$]{\label{d3}\includegraphics[width=3.9cm,height=4cm]{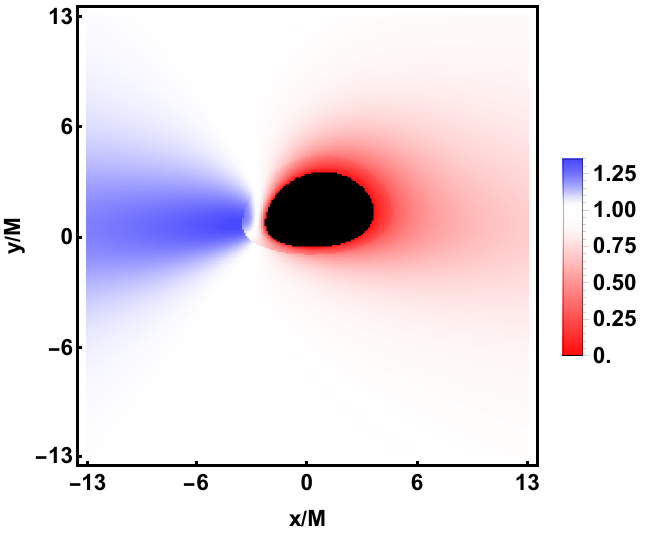}}
\caption{The red-shift factors of direct images of rotating BHs in
Horndeski gravity with different values of $a$ and $h$ with fixed
$\theta_{obs}=70^{\circ}$ under prograde flow. The red and blue
colours represent the red-shift and blue-shift, respectively, while
the solid black region depict the inner shadows.}\label{prd6}
\end{center}
\end{figure}

We now proceed to investigate the imaging process of the BH in
greater detail and present an accurate evaluation of the red-shift
factors associated with the motion of emitting particles. These
results can be interpreted through the framework of the Doppler
effect. In Fig. \textbf{\ref{prd6}}, we discuss the impact of
relevant parameters on the distribution of red-shift factors. This
include the direct images of the prograde flows, where red and blue
colours corresponds to red-shift and blue-shift, respectively. In
the figures, the black region at the center of the image indicates
the inner shadows of the BH. Importantly, in all cases, the
blue-shift appearing on the left side of the screen, while the
red-shift are on the right side of the screen. Here, we observe that
the variation of rotation parameter $a$ (see each column from top to
bottom) has significantly impact on the red-shift factors, such as
the observational intensity of red-shift factors are decreases with
the increasing values of $a$. On the other hand, increasing the hair
parameter $h$ (see each row from left to right) leads to a slight
decrease in the red-shift luminosity and occupies less space on the
screen as compared to smaller values. Additionally, in all these
images, the red-shift factor forms a strict boundary around the
solid black region, while the blue-shift region is positioned
slightly away from the red-shift boundary. We calculated the maximal
blue-shift $g_{max}$ of direct images under different values of $a$
and $h$ with $\theta_{obs}=70^{\circ}$ in Table \textbf{\ref{tab1}}.
From this table, it can be observed that the numerical values of
$g_{max}$ decrease with increasing $a$, while they increase with
larger values of $h$.

\begin{figure}[H]
\begin{center}
\subfigure[\tiny][~$a=0.001,~h=0.001$]{\label{a1}\includegraphics[width=3.9cm,height=4cm]{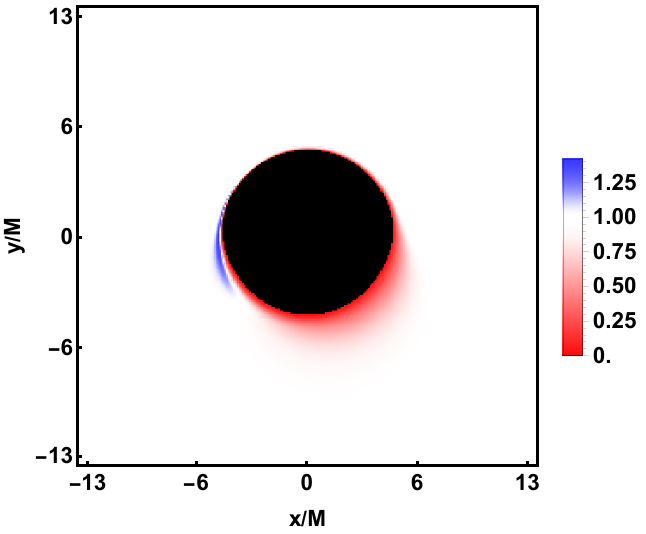}}
\subfigure[\tiny][~$a=0.001,~h=0.1$]{\label{b1}\includegraphics[width=3.9cm,height=4cm]{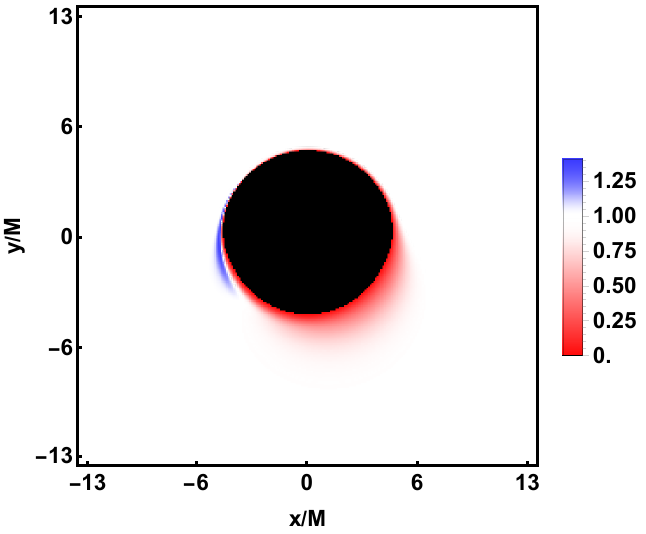}}
\subfigure[\tiny][~$a=0.001,~h=0.5$]{\label{c1}\includegraphics[width=3.9cm,height=4cm]{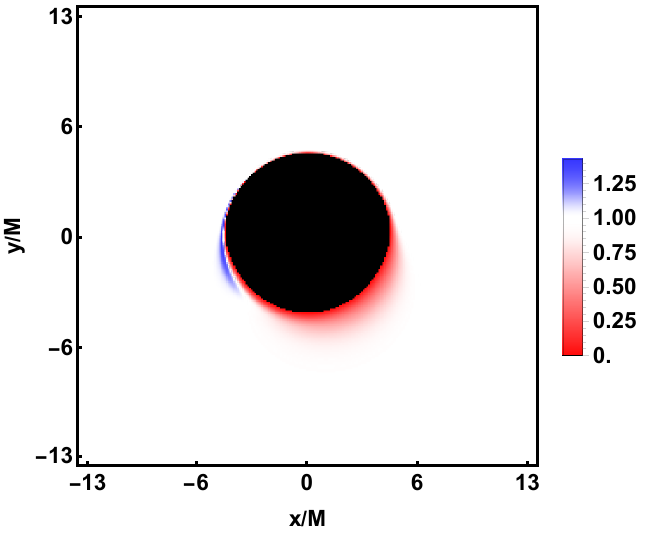}}
\subfigure[\tiny][~$a=0.001,~h=0.9$]{\label{d1}\includegraphics[width=3.9cm,height=4cm]{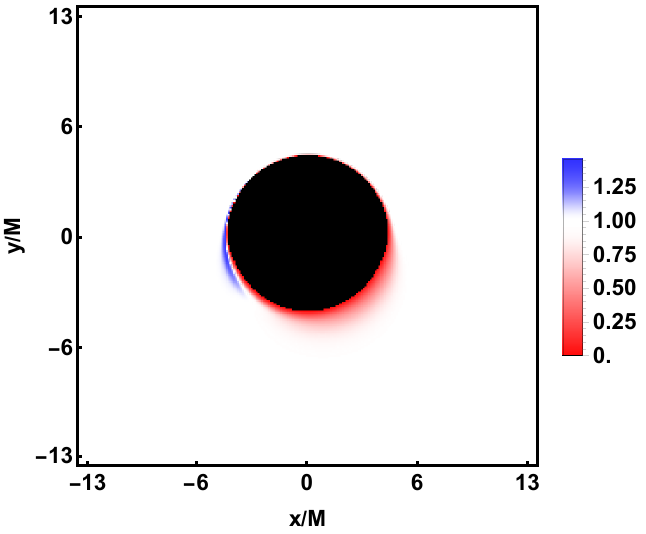}}
\subfigure[\tiny][~$a=0.5,~h=0.001$]{\label{a2}\includegraphics[width=3.9cm,height=4cm]{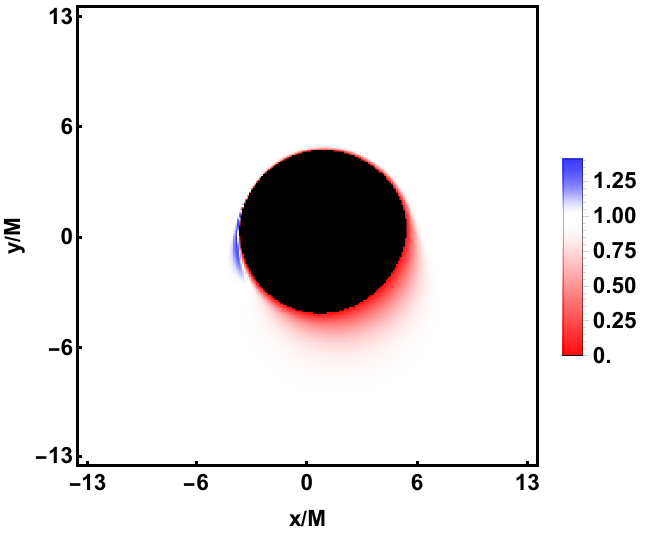}}
\subfigure[\tiny][~$a=0.5,~h=0.1$]{\label{b2}\includegraphics[width=3.9cm,height=4cm]{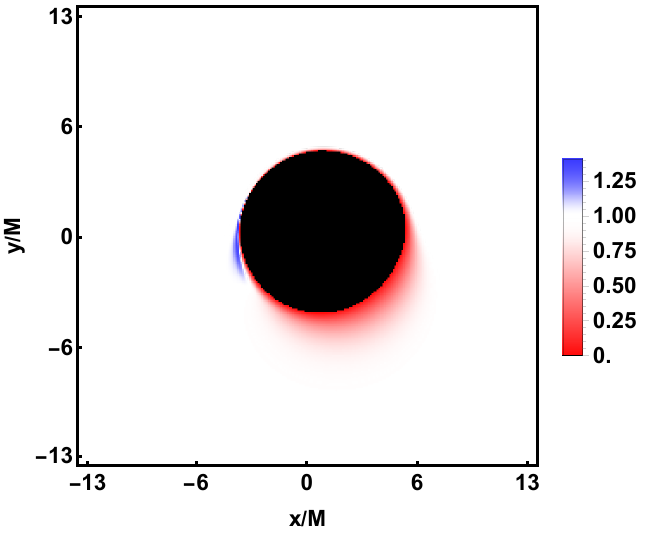}}
\subfigure[\tiny][~$a=0.5,~h=0.5$]{\label{d2}\includegraphics[width=3.9cm,height=4cm]{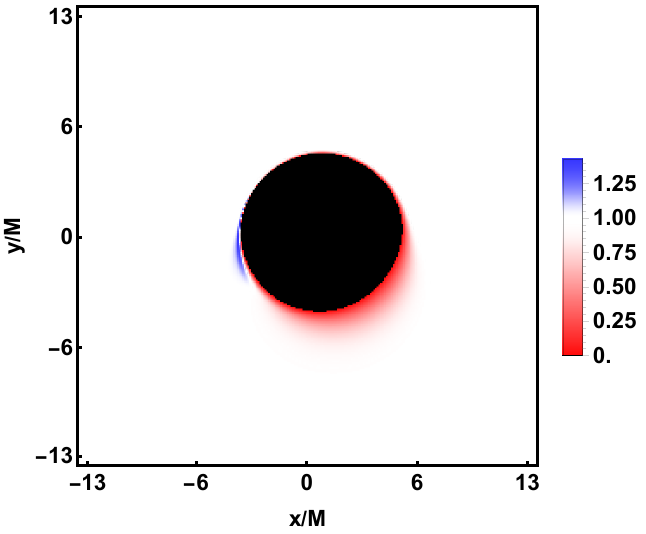}}
\subfigure[\tiny][~$a=0.5,~h=0.9$]{\label{d2}\includegraphics[width=3.9cm,height=4cm]{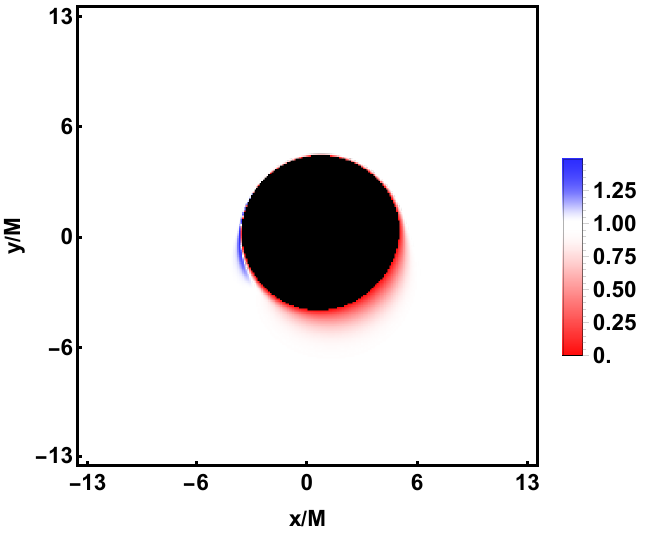}}
\subfigure[\tiny][~$a=0.9,~h=0.001$]{\label{a3}\includegraphics[width=3.9cm,height=4cm]{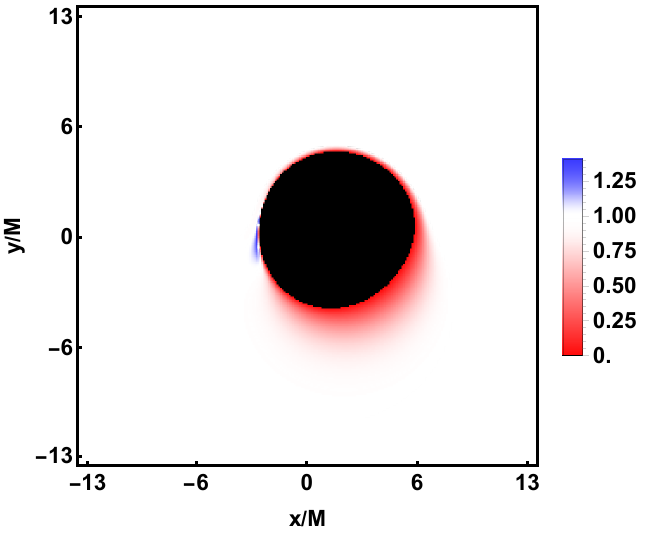}}
\subfigure[\tiny][~$a=0.9,~h=0.1$]{\label{b3}\includegraphics[width=3.9cm,height=4cm]{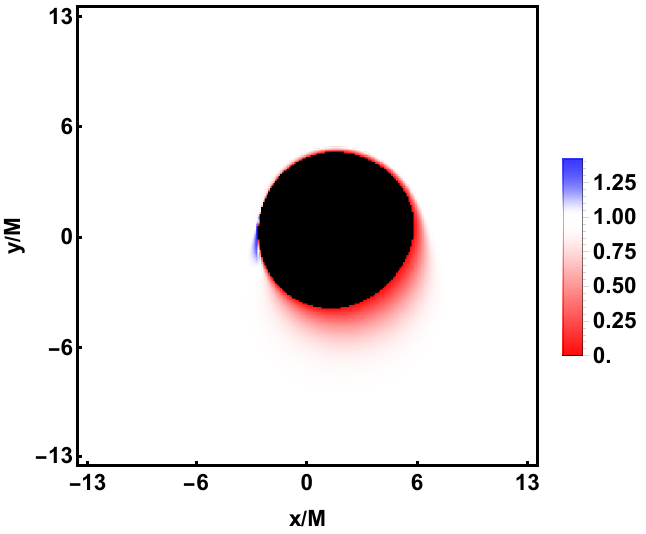}}
\subfigure[\tiny][~$a=0.9,~h=0.5$]{\label{c3}\includegraphics[width=3.9cm,height=4cm]{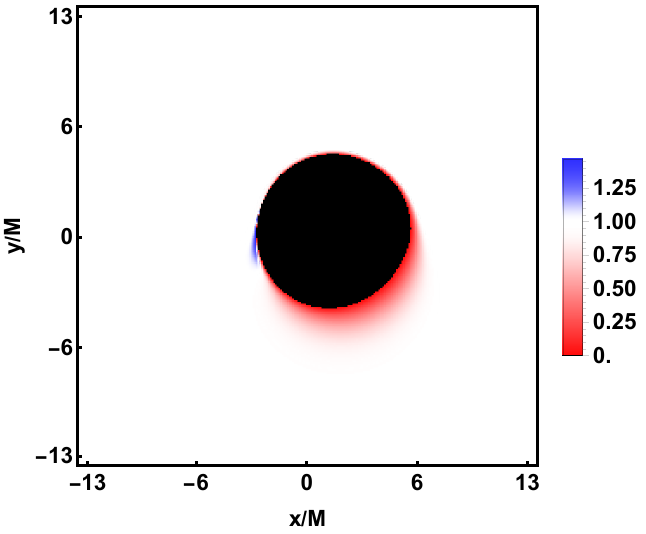}}
\subfigure[\tiny][~$a=0.9,~h=0.9$]{\label{d3}\includegraphics[width=3.9cm,height=4cm]{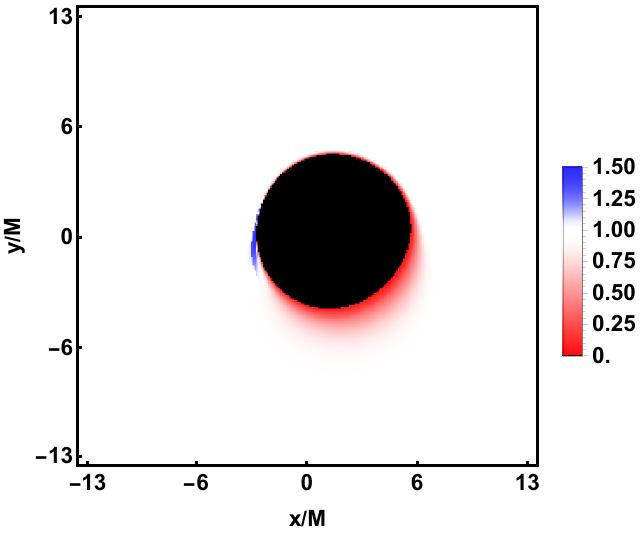}}
\caption{The red-shift factors of lensed images of rotating BHs in
Horndeski gravity with different values of $a$ and $h$ with fixed
$\theta_{obs}=70^{\circ}$ under prograde flow. The red and blue
colours represent the red-shift and blue-shift, respectively, while
the solid black region depict the inner shadows.}\label{prd7}
\end{center}
\end{figure}

The optical appearance of the red-shift distribution for the lensed
images with prograde flow is illustrated in Fig.
\textbf{\ref{prd7}}, where the parameter values correspond to those
in Fig. \textbf{\ref{prd6}}. In all cases, one can noticed that the
outer edge of inner shadow is enveloped by a strict red
crescent-like shape and the optical appearance of the red-shift
colour is expand in the lower right quadrant of the screen. On the
other hand, the blue-shift factor are appeared a small petal like
shape on the left side of the screen. Consequently, as the
parameters vary, the lensed images of the accretion disk
predominantly exhibit red-shift features, while the blue-shift
factor is significantly suppressed. Notably, with the enhancement of
the both $a$ and $h$, the observational appearance of blue-shift
factor is notably diminished. Generally, the rotation parameter $a$
and the hair parameter $h$ have a stronger influence on the
red-shift configuration in the lensed images as compared to their
impact on the direct images.

\begin{table}[H]\centering
\begin{tabular}{|c|c|c|c|c|c|c|}
\hline \diagbox{$h$}{$a$} & 0.001 & 0.1 & 0.3 & 0.5 & 0.7 & 0.9 \\
\hline 0.001 & 3.12947 & 2.54799 & 1.91835 & 1.59067 & 1.3773 &
1.31087  \\ \hline 0.1 & 3.24286 & 2.63024 & 1.96354 & 1.62373 &
1.40265 & 1.31885  \\ \hline 0.3 & 3.59302 & 2.80229 & 2.07119 &
1.70615 & 1.46472 & 1.33552 \\ \hline 0.5 & 4.01729 & 3.09129 &
2.19797 & 1.78525 & 1.52903 & 1.35339  \\ \hline
0.7 & 4.36971 & 3.37750 & 2.34050 & 1.86013 & 1.59435 & 1.37149  \\
\hline 0.9 & 5.08319 & 3.64835 & 2.47521 & 1.94328 & 1.66057 &
1.44162 \\ \hline
\end{tabular}
\caption{The maximal blue-shift $g_{max}$ of direct images under
different values of $a$ and $h$ with
$\theta_{obs}=70^{\circ}$.}\label{tab1}
\end{table}

\begin{figure}[H]
\begin{center}
\subfigure[\tiny][~$a=0.001,~h=0.001$]{\label{a1}\includegraphics[width=3.9cm,height=4cm]{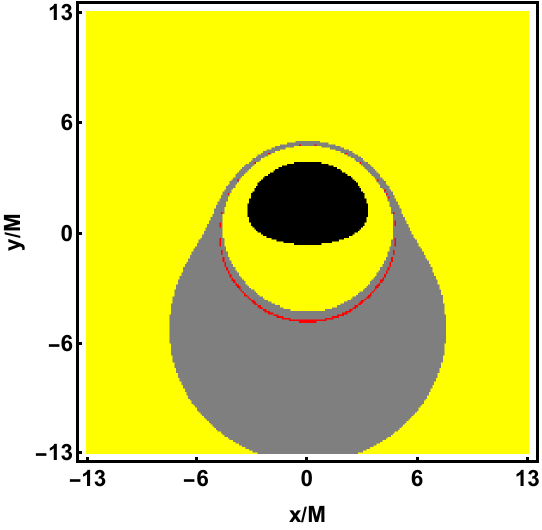}}
\subfigure[\tiny][~$a=0.001,~h=0.1$]{\label{b1}\includegraphics[width=3.9cm,height=4cm]{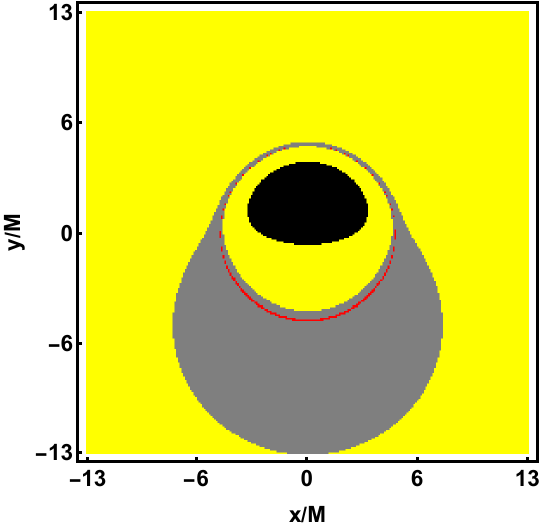}}
\subfigure[\tiny][~$a=0.001,~h=0.5$]{\label{c1}\includegraphics[width=3.9cm,height=4cm]{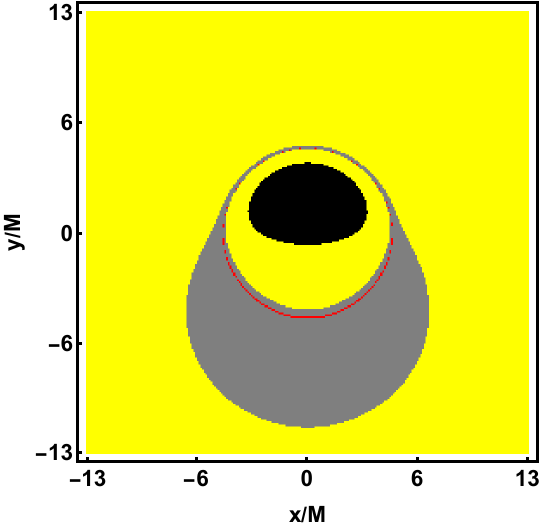}}
\subfigure[\tiny][~$a=0.001,~h=0.9$]{\label{d1}\includegraphics[width=3.9cm,height=4cm]{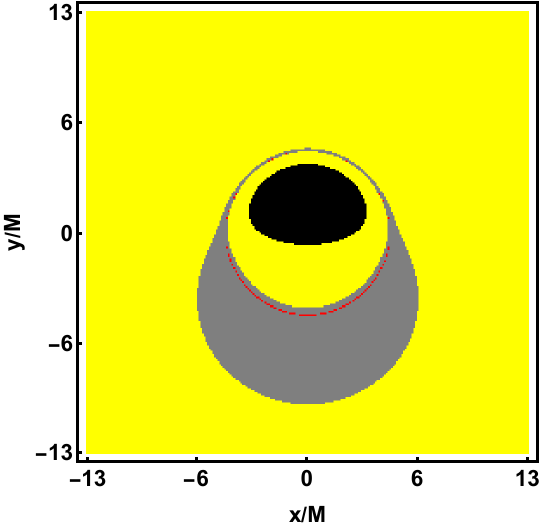}}
\subfigure[\tiny][~$a=0.5,~h=0.001$]{\label{a2}\includegraphics[width=3.9cm,height=4cm]{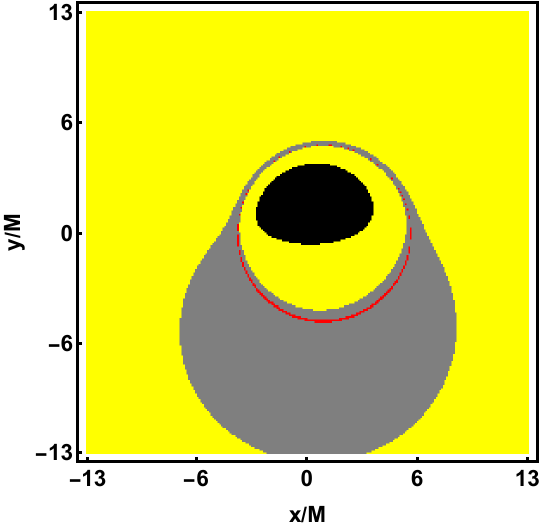}}
\subfigure[\tiny][~$a=0.5,~h=0.1$]{\label{b2}\includegraphics[width=3.9cm,height=4cm]{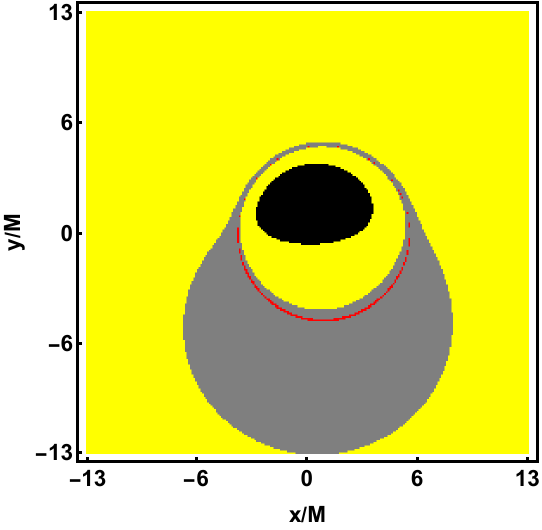}}
\subfigure[\tiny][~$a=0.5,~h=0.5$]{\label{d2}\includegraphics[width=3.9cm,height=4cm]{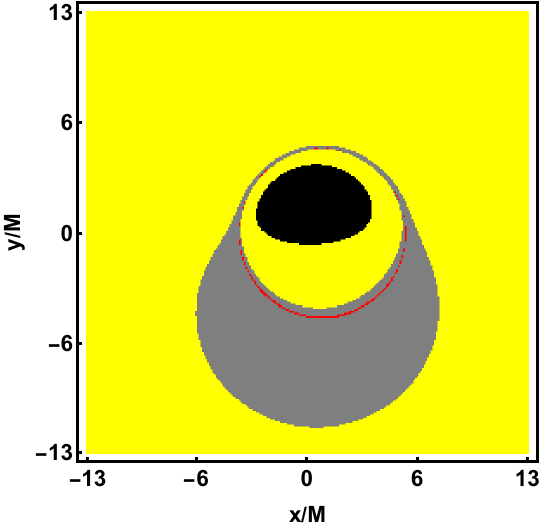}}
\subfigure[\tiny][~$a=0.5,~h=0.9$]{\label{d2}\includegraphics[width=3.9cm,height=4cm]{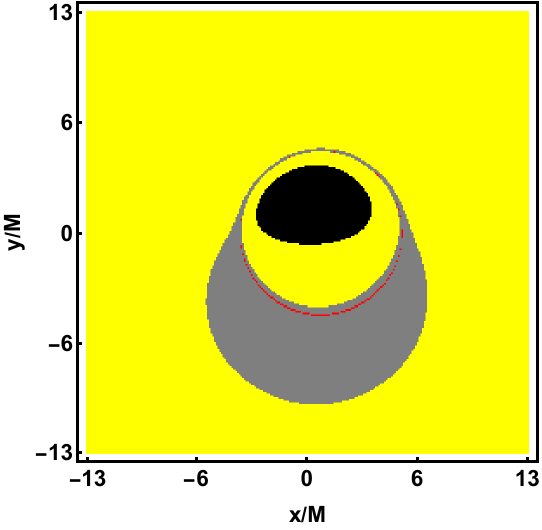}}
\subfigure[\tiny][~$a=0.9,~h=0.001$]{\label{a3}\includegraphics[width=3.9cm,height=4cm]{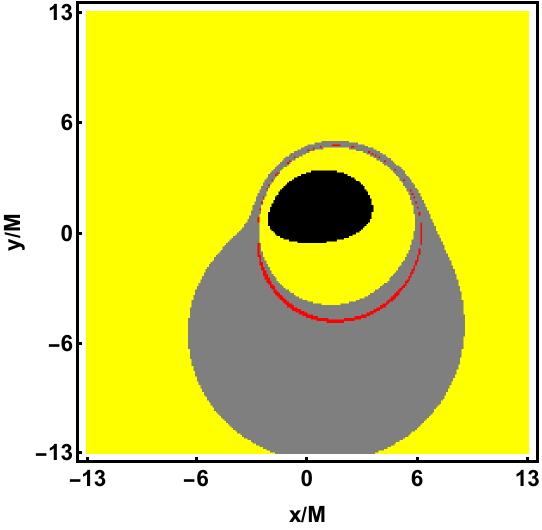}}
\subfigure[\tiny][~$a=0.9,~h=0.1$]{\label{b3}\includegraphics[width=3.9cm,height=4cm]{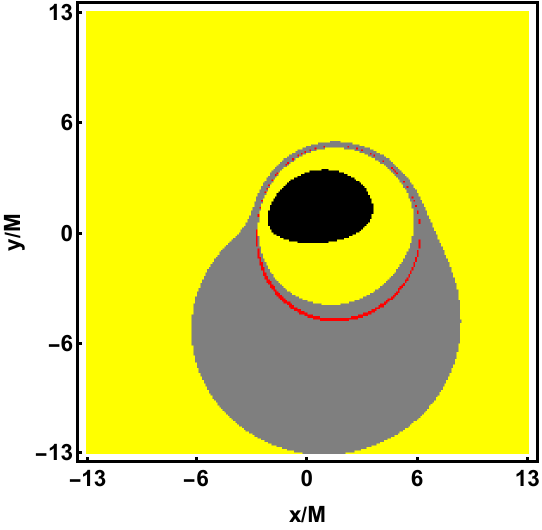}}
\subfigure[\tiny][~$a=0.9,~h=0.5$]{\label{c3}\includegraphics[width=3.9cm,height=4cm]{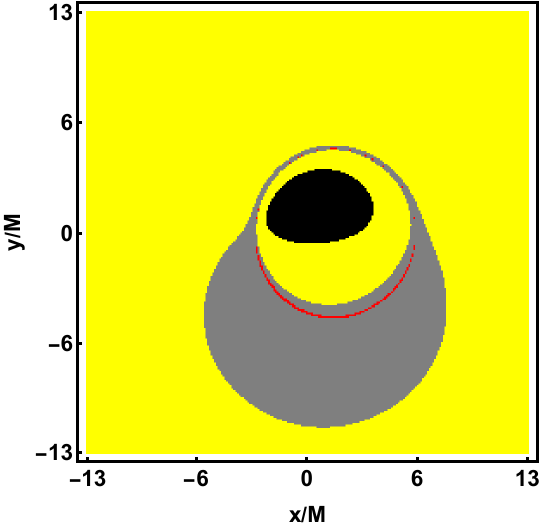}}
\subfigure[\tiny][~$a=0.9,~h=0.9$]{\label{d3}\includegraphics[width=3.9cm,height=4cm]{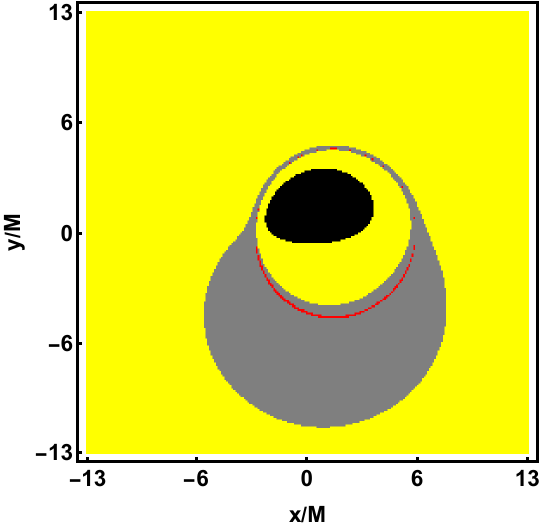}}
\caption{The lensing bands of rotating BHs in Horndeski gravity with
different values of $a$ and $h$ with fixed $\theta_{obs}=70^{\circ}$
under prograde flow. The colours yellow, grey and red correspond to
the direct, lensed and photon ring images, respectively, while the
solid black region depict the inner shadows.}\label{prd8}
\end{center}
\end{figure}

\begin{figure}[H]
\begin{center}
\subfigure[\tiny][~$h=0.001$]{\label{a1}\includegraphics[width=3.9cm,height=3.8cm]{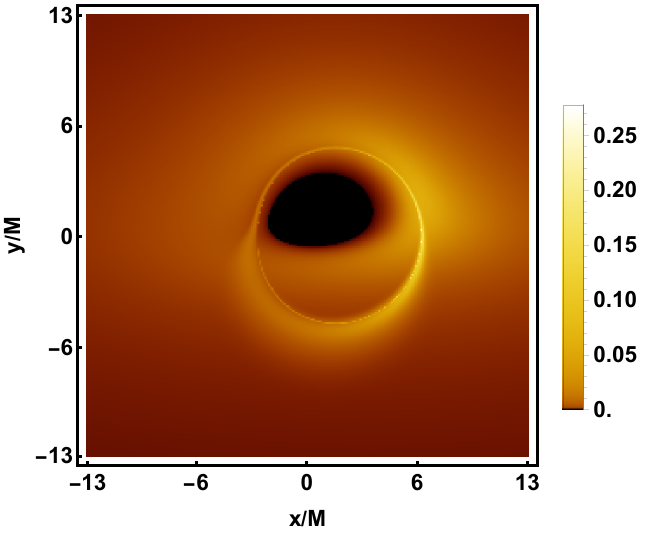}}
\subfigure[\tiny][~$h=0.1$]{\label{b1}\includegraphics[width=3.9cm,height=3.8cm]{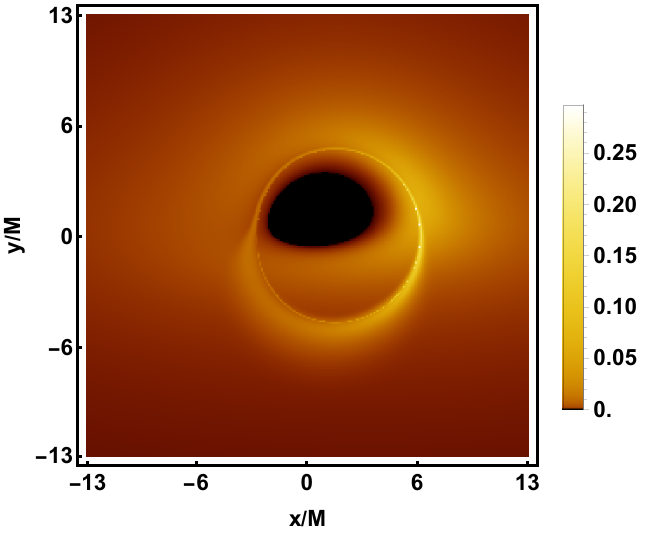}}
\subfigure[\tiny][~$h=0.5$]{\label{c1}\includegraphics[width=3.9cm,height=3.8cm]{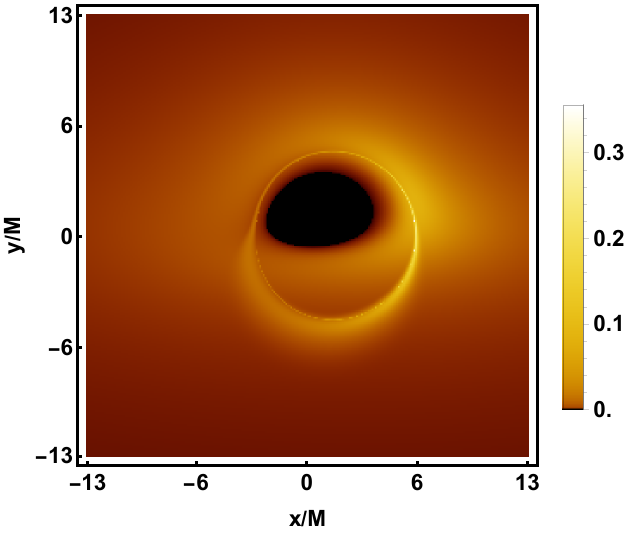}}
\subfigure[\tiny][~$h=0.9$]{\label{d1}\includegraphics[width=3.9cm,height=3.8cm]{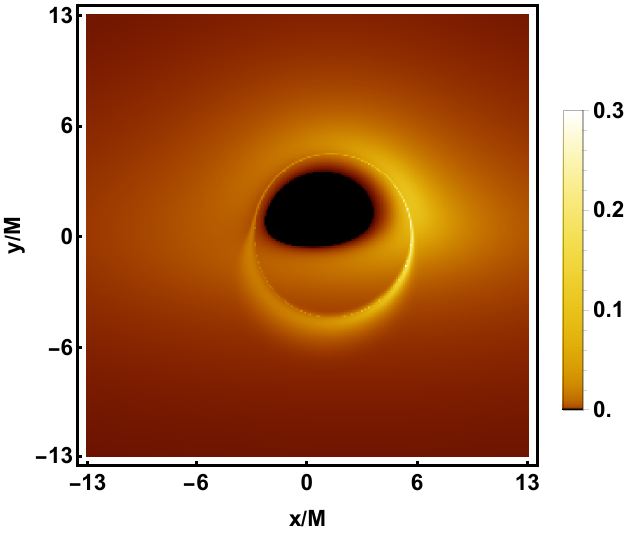}}
\caption{Optical images of rotating BHs in Horndeski gravity for
different values of $h$ with fixed $a=0.9$ and
$\theta_{obs}=70^{\circ}$ under retrograde flow. The BH's event
horizon is represented as a black region and a luminous circular
ring corresponds to the position of the bright photon
ring.}\label{prd9}
\end{center}
\end{figure}

For the differentiation between direct and lensed images of the
accretion disk, we exhibit the corresponding observed fluxes in Fig.
\textbf{\ref{prd8}}. In this figures, the lensing bands appeared
with three different colours such as yellow, grey and red, which is
correspond to the direct, lensed and photon ring images,
respectively, while the solid black region depict the inner shadows.
Upon comparing, from the first row of Fig. \textbf{\ref{prd8}}, we
noticed that the lensed bands always appear in the lower half
quadrant of the screen, and significantly shrink towards the upper
side of the screen with the increasing values of $h$. When $a=0.5$
(see second row of Fig. \textbf{\ref{prd8}}) alterations in $h$
significantly deformed the lensed bands in the lower half of the
screen and slightly moves towards the right side. Moreover, the
corresponding radius of photon ring are decreases with the aid of
$h$. Similarly, when we further increase the value of $a$, such that
$a=0.9$ (see third row of Fig. \textbf{\ref{prd8}}), the lensed band
significantly deformed and moves towards the right side of the
screen. Moreover, with the increasing values of $h$, the lensed
bands are shrink towards the upper side of the screen and the
corresponding radius of photon ring are decreases. Notably, in all
cases, the photon ring consistently lies entirely within the yellow
and grey bands, while the inner shadow maintains an almost constant,
hat-like shape.
Now, we discuss the impact of the hair parameter $h$ on the visual
characteristics of the rotating BHs in Horndeski gravity under the
retrograde thin accretion disk flow. In contrast to the prograde
case, the gravitational red-shift effect markedly diminishes the
observed brightness of the shadow image. The overall reduction in
light intensity makes it challenging to distinguish the lensed image
from higher-order images, which leads to slightly decreasing the
clarity of the photon ring. Looking at Fig. \textbf{\ref{prd9}}
(a-d), we observed that in the lower right half quadrant of the
screen there is a luminous matter, which is slightly shrink towards
the upper region of the screen with the increasing values of $h$.
Moreover, in the upper right side of the screen a
``crescent-shaped'' bright region appears, which is slightly enhance
with the aid of $h$. This is because the jet material close the side
of the event horizon interpret low brightness, while during the
imaging process, the radiating material confined to the equatorial
plane increases the total optical depth.

\begin{figure}[H]
\begin{center}
\subfigure[\tiny][~$h=0.001$]{\label{a1}\includegraphics[width=3.9cm,height=3.8cm]{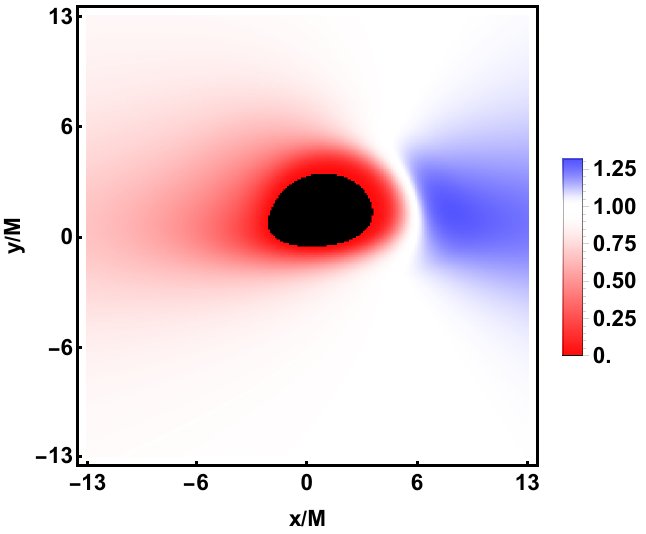}}
\subfigure[\tiny][~$h=0.1$]{\label{b1}\includegraphics[width=3.9cm,height=3.8cm]{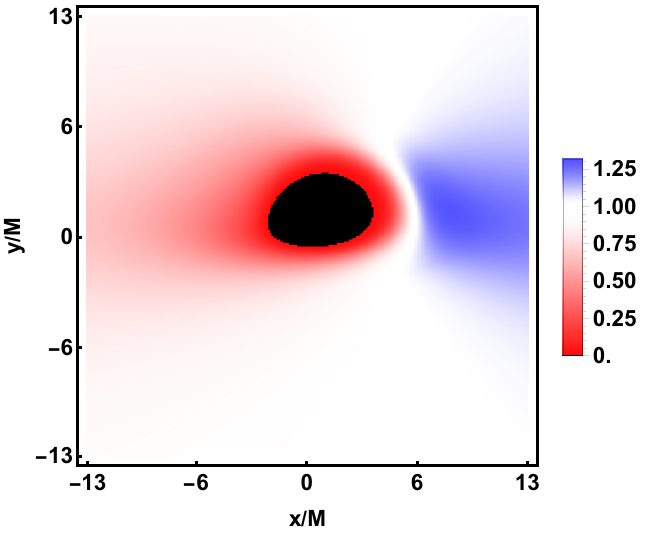}}
\subfigure[\tiny][~$h=0.5$]{\label{c1}\includegraphics[width=3.9cm,height=3.8cm]{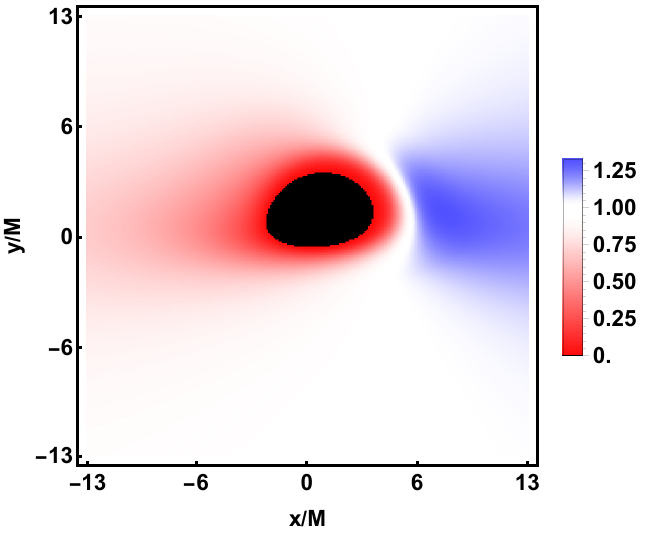}}
\subfigure[\tiny][~$h=0.9$]{\label{d1}\includegraphics[width=3.9cm,height=3.8cm]{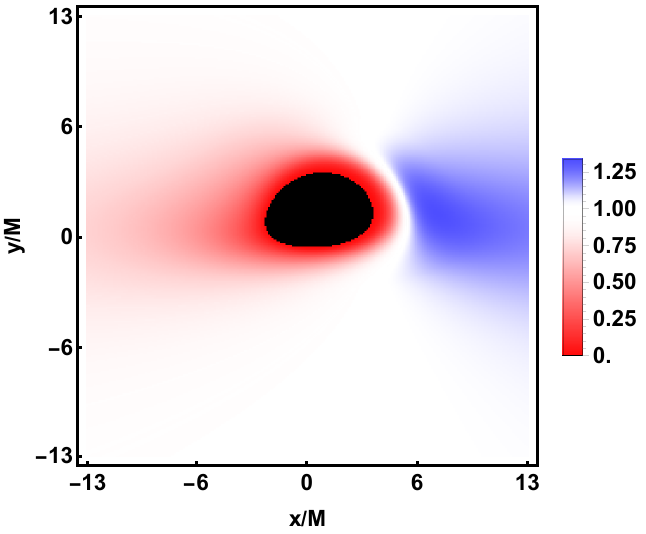}}
\caption{ The red-shift factors of direct images of rotating BHs in
Horndeski gravity for different values of $h$ with fixed $a=0.9$ and
$\theta_{obs}=70^{\circ}$ under retrograde flow. The red and blue
colours represent the red-shift and blue-shift, respectively, while
the solid black region depict the inner shadows.}\label{prd10}
\end{center}
\end{figure}

\begin{figure}[H]
\begin{center}
\subfigure[\tiny][~$h=0.001$]{\label{a1}\includegraphics[width=3.9cm,height=3.8cm]{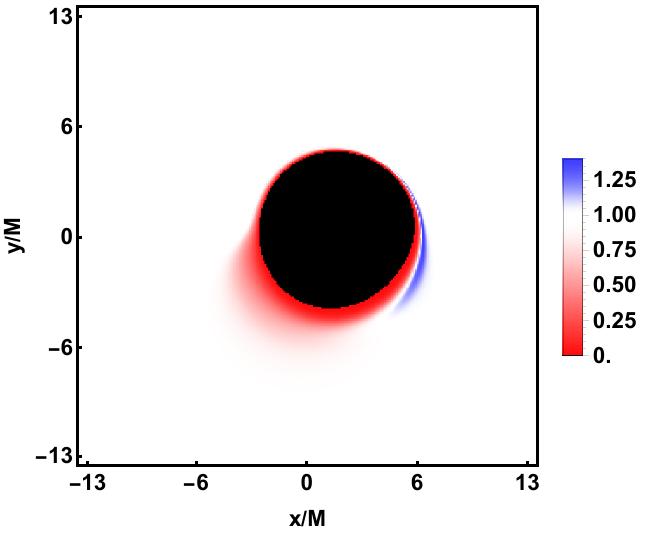}}
\subfigure[\tiny][~$h=0.1$]{\label{b1}\includegraphics[width=3.9cm,height=3.8cm]{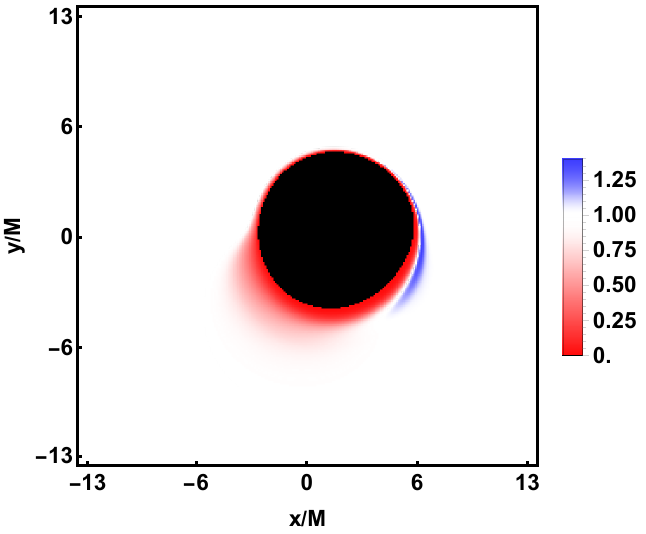}}
\subfigure[\tiny][~$h=0.5$]{\label{c1}\includegraphics[width=3.9cm,height=3.8cm]{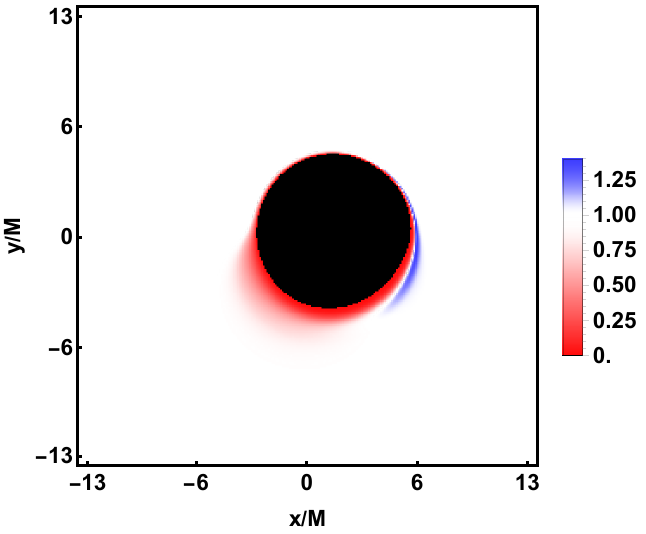}}
\subfigure[\tiny][~$h=0.9$]{\label{d1}\includegraphics[width=3.9cm,height=3.8cm]{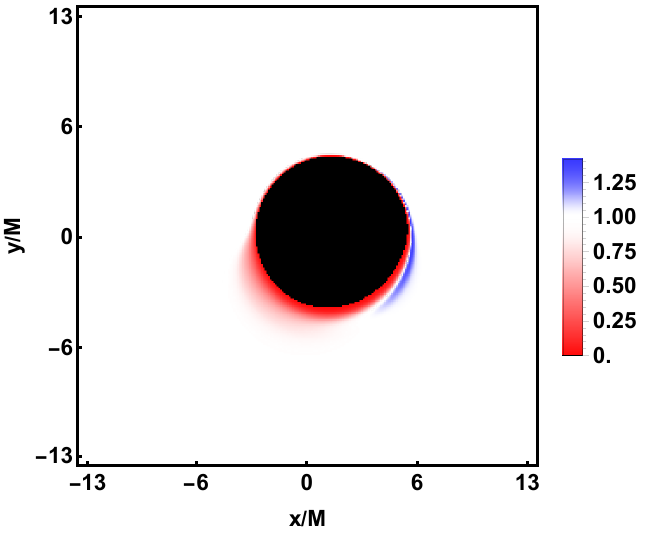}}
\caption{ The red-shift factors of lensed images of rotating BHs in
Horndeski gravity for different values of $h$ with fixed $a=0.9$ and
$\theta_{obs}=70^{\circ}$ under retrograde flow. The red and blue
colours represent the red-shift and blue-shift, respectively, while
the solid black region depict the inner shadows.}\label{prd11}
\end{center}
\end{figure}

\begin{figure}[H]
\begin{center}
\subfigure[\tiny][~$h=0.001$]{\label{a1}\includegraphics[width=3.9cm,height=3.8cm]{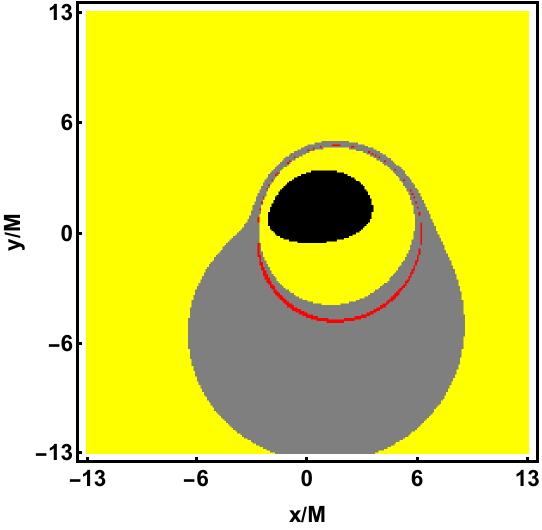}}
\subfigure[\tiny][~$h=0.1$]{\label{b1}\includegraphics[width=3.9cm,height=3.8cm]{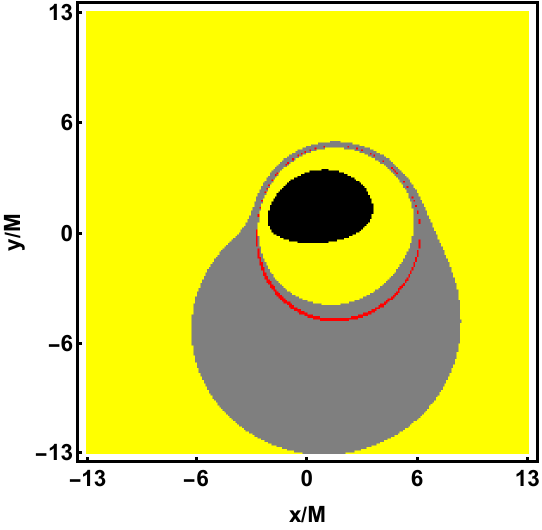}}
\subfigure[\tiny][~$h=0.5$]{\label{c1}\includegraphics[width=3.9cm,height=3.8cm]{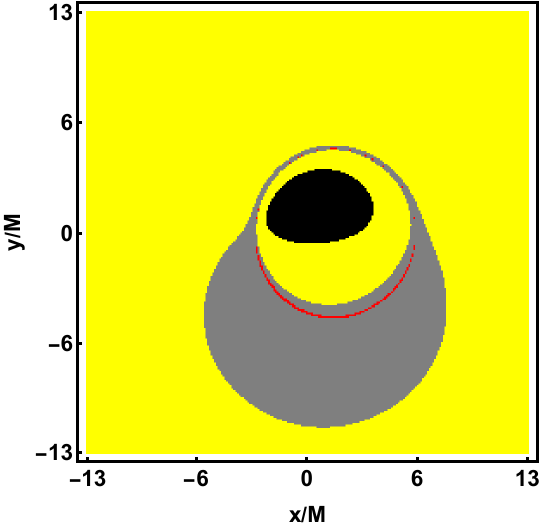}}
\subfigure[\tiny][~$h=0.9$]{\label{d1}\includegraphics[width=3.9cm,height=3.8cm]{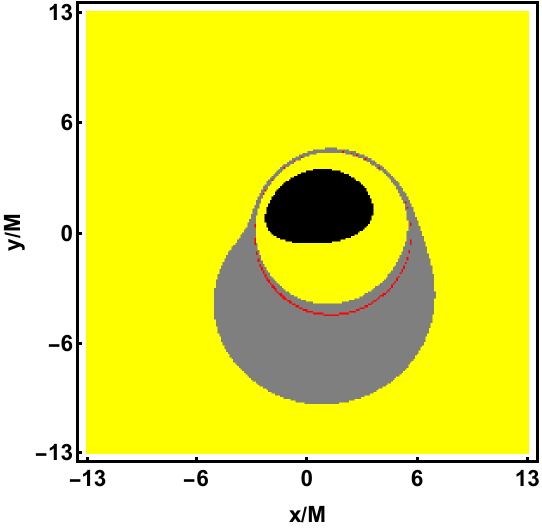}}
\caption{The lensing bands of rotating BHs in Horndeski gravity for
different values of $h$ with fixed $a=0.9$ and
$\theta_{obs}=70^{\circ}$ under retrograde flow. The colours yellow,
grey and red correspond to the direct, lensed and photon ring
images, respectively, while the solid black region depict the inner
shadows.}\label{prd12}
\end{center}
\end{figure}

In Fig. \textbf{\ref{prd10}}, we exhibit the visual characteristics
of red-shift distributions of direct images under retrograde flow
for different values of $h$. Upon comparing with the prograde flow,
in this case the red and blue colours are in opposite direction.
From these images, one can find that the phenomenon of red-shift
lensing expands in more space and envelops the inner shadow of BH
more obviously. Importantly, in all images the impact of $h$ on the
distributions of red-shift factors are negligible. For the lensed
images, we observe that the red-shift distribution appears only in
the lower-left quadrant of the screen, with its strip gradually
decreases with the increasing of $h$ (see Fig.
\textbf{\ref{prd11}}). On the right side of the screen, a narrow
strip of blue-shift appears near the inner shadow, remaining almost
unaffected with the variations of $h$. In Fig. \textbf{\ref{prd12}},
we exhibit the optical images of lensing bands of the considering BH
model for different values of $h$ with retrograde accretion flow.
From these images, we noticed that the lensed images are gradually
shrink towards the lower right side of the screen with the
increasing values of $h$. Additionally, the inner shadow interprets
a hat-like shape, which remains almost constant with the variations
of $h$ and photon ring always stay within the confines of direct and
lensed images.

\section{Constraints on Parameters with the EHT Observations}

\begin{figure}[H]
\begin{center}
\subfigure[\tiny][~Shadow angular diameter of
M$87^{\ast}$]{\label{a1}\includegraphics[width=8cm,height=5.5cm]{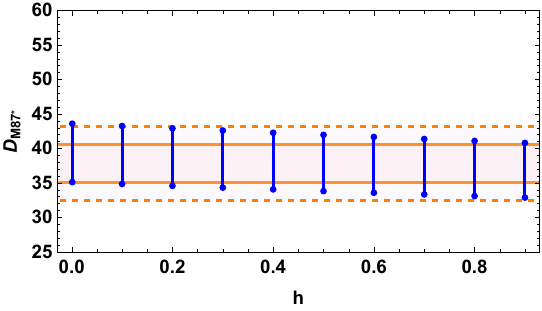}}
\subfigure[\tiny][~Shadow angular diameter of Sgr
$A^{\ast}$]{\label{b1}\includegraphics[width=8cm,height=5.5cm]{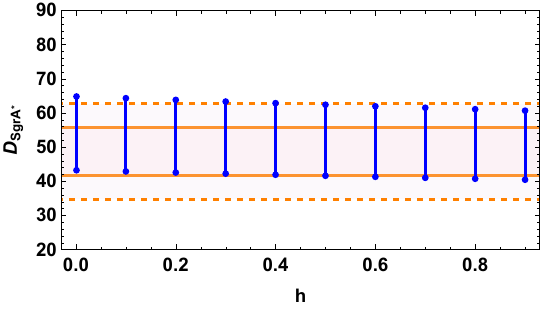}}
\caption{Plots showing the approximated shadow angular diameter $D$.
The solid and dashed orange lines indicate the $1\sigma$ and
$2\sigma$ confidence levels for $D$, respectively, whereas the blue
strip bars is the estimated ranges. In both cases, we fixed
$a=0.5$.}\label{israr1}
\end{center}
\end{figure}

In this subsection, we find the constraints to the hair parameter
$h$ using the observation data provided by the EHT for M$87^{\ast}$
and Sgr $A^{\ast}$. In order to discuss the desired results, we
determine shadow angular diameter $D$ and made a comparative
analysis with the angular diameter of M$87^{\ast}$ and Sgr
$A^{\ast}$. For this, the angular diameter of the BH is expressed as
$D=2\mathcal{R}_{d}\frac{\mathcal{M}}{\mathcal{D}_{obs}}$, where
$\mathcal{R}_{d}$ represents the radius of BH shadow when the
observer's frame lies at the BH position, which is related to
$R_{d}$ and $\mathcal{M}$ is the BH's mass lies at distance
$\mathcal{D}_{obs}$ from the observer \cite{en30,sd65}.
Mathematically, the angular diameter can be expressed as
\cite{en30,sd65}
\begin{equation}\label{eht1}
D=2\times9.87098\mathcal{R}_{d}\big(\frac{\mathcal{M}}{M_{\odot}}\big)\big(\frac{1\text{kpc}}{\mathcal{D}_{obs}}\big)\mu
as.
\end{equation}
Using Eq. (\ref{eht1}), one can determine the theoretical angular
diameters for different parameter configurations and compare them
with actual astronomical observational bounds. For M$87^{\ast}$, the
distance from Earth is $\mathcal{D}_{obs}=16.8\text{kpc}$ and the
estimated BH mass is $\mathcal{M}=(6.5\pm0.7)\times
10^{6}M_{\odot}$, whereas the shadow diameter is
$D_{M87^{\ast}}=(37.8\pm2.7)\mu as$ \cite{sd66}. For Sgr $A^{\ast}$,
its distance from Earth is about $\mathcal{D}_{obs}=8\text{kpc}$ and
its estimated BH mass is $\mathcal{M}=(4.0^{+1.1}_{-0.6})\times
10^{6}M_{\odot}$, whereas the shadow diameter is $D_{Sgr
A^{\ast}}=(48.7\pm7)\mu as$ \cite{sd67}. In this review, we have
presented the shadow angular diameter $D$ with respect to the hair
parameter $h$ for M$87^{\ast}$ and Sgr $A^{\ast}$ in left and right
panels of Fig. \textbf{\ref{israr1}}, respectively with fixed
$a=0.5$. From this this figure, one can notice that the estimated
interval remains almost within $1\sigma$ and $2\sigma$ confidence
constraints. This indicates that, the constraints on $h$ provided by
both M$87^{\ast}$ and Sgr $A^{\ast}$ are satisfied nicely. Hence,
the astronomical observations impose strong constraints on the hair
parameter $h$ for both M$87^{\ast}$ and Sgr $A^{\ast}$. Moreover,
since the rotation parameter $a$ only affects the deviation from the
circularity of the shadow and has minor impact on its radius, it
will not be discussed here. More comprehensive and precise future
observations of BHs will be essential for further refining the
constraints on the hair parameter $h$.

\section{Conclusion}
Recently, the study of BH shadow images is a hot topic in both
astrophysics and theoretical physics. The novel developments of the
EHT have opened new pathways for utilizing BH images in the study of
high energy physics and gravitational theories, sparking significant
interest in numerical simulations of BH images across the scientific
community. In the imaging process of BHs, the accretion of matter
surrounded by high-energy radiation plays a important role. In this
study, we consider two different background light sources such as
the celestial light source and the thin accretion disk. To examine
their visual characteristics, we consider the rotating BH within the
framework of Horndeski gravity and determine how varies in
associated parameters influence the visual characteristics of its
image. Our results indicate that in the analysis of BH horizon
$\Delta(r)$, the horizon radii depend on the both $h$ and $a$, as
horizon radii decreases with the increasing of $a$. While the gap
between the curves are slightly increases with the increasing of
$h$. The shadow contours reveals that, an decrease in $h$, causes
the shadow contours shift significantly towards the right side of
the screen. Moreover, when $h$ has maximum value such as $h=3.1$ the
shadow contour to transition from a circular configuration to a
sightly ``D'' shape, reflecting a pronounced frame-dragging effect.

Based on celestial light source model, we have analysed the visual
characteristics of BHs under different values of relevant
parameters. These results illustrates that the shape and size of the
BH shadow image are related to alteration in the rotation parameter
$a$ and the hair parameter $h$. Generally, it is noticed that as the
values of $h$ are increases, the ``D'' shape petals are slightly
evolves and the radius of the white circular ring are gradually
moves towards the interior of the BH. On the other hand, the larger
values of $a$ leads to increase the space-dragging effect and the
optical images of inner shadow are slightly deform into ``D'' shape
images. Consequently, in this model, differences in the impact of
these two parameters on the optical images of the rotating BH in
Horndeski gravity can be clearly distinguished. For thin accretion
disk, we have analysed the significant features of accreting matter
surrounding the BH for two different accretion flow such as prograde
and retrograde. In the case of prograde accretion flow, it is
observed that with the variations of $h$, the shadow images are
slightly deformed, emerging a smooth, hat-like black region. On the
other hand, with the variations of $a$ (see Fig. \textbf{\ref{prd5}}
from top to bottom), the size of the inner shadow are significantly
decreases and a ``crescent-shaped'' bright region appears on the
left side of the critical curve, significantly increases the
corresponding intensity with the larger values of $a$. We have
discussed the red-shift configuration of accreting matter around BH
for both direct and lensed images in Figs. \textbf{\ref{prd6}} and
\textbf{\ref{prd7}}, respectively. In the case direct images, the
blue-shift appearing on the left side of the screen, while the
red-shift are on the right side of the screen. We observed that the
variation of rotation parameter $a$ has significantly impact on the
red-shift factors, such as the observational intensity of red-shift
factors are decreases with the increasing values of $a$. Moreover,
the increasing values of the hair parameter $h$ leads to a slight
decrease in the red-shift luminosity and occupies less space on the
screen as compared to smaller values. A prominent features is
observed, as the blue-shift maps appear significantly smaller than
the red-shift maps. For lensed images, the exterior boundary of
inner shadow is enveloped by a strict red crescent-like shape and
the optical appearance of the red-shift colour is expand in the
lower right quadrant of the screen. And the blue-shift factor are
appeared a small petal like shape on the left side of the screen.
The variations in parameters, results in suppressed the blue-shift
factor as compared to red-shift. In Fig. \textbf{\ref{prd8}}, we
interpret the lensing bands of accretion matter under different
values of $a$ and $h$. The results indicate that with the
enlargement of $a$, the lensed bands (see grey colour) gradually
expand and significantly deform towards the lower right quadrant of
the screen. On the other hand, the increasing values of $h$ leads to
shrink the lensed bands towards the upper side of the screen and the
corresponding radius of photon ring are gradually decreases.

Subsequently, we examine the optical images of the rotating BH
model, considering the case where the accretion flow is retrograde.
It is found that in the lower right half quadrant of the screen
there is a luminous matter, which is slightly shrink towards the
upper region of the screen with the increasing values of $h$ (see
Fig. \textbf{\ref{prd9}}). And in the upper right side of the screen
a ``crescent-shaped'' bright region exhibit, which is slightly
enhance with the variations of $h$. We also investigate the
red-shift configuration for both direct and lensed images in Figs.
\textbf{\ref{prd10}} and \textbf{\ref{prd11}}, respectively. From
these figures, we noticed that the phenomenon of red-shift lensing
expands in more space and envelops the inner shadow of BH more
obviously and the impact of $h$ on the distributions of red-shift
factors are hardly differentiable. For the lensed images, we
observed that the red-shift distribution appears only in the
lower-left quadrant of the screen, and on the right side of the
screen, a narrow strip of blue-shift appears near the inner shadow,
remaining almost constant with the variations of $h$. In Fig.
\textbf{\ref{prd12}}, it is observed that as $h$ increases, the
lensed band gradually shrinks toward the lower-right region of the
screen, while the photon ring consistently remains confined between
the direct and lensed images. Finally, we consider the recent
observational data from M$87^{\ast}$ and Sgr $A^{\ast}$ to impose
certain parameter constraints on rotating BHs, confirming the
validity of Horndeski gravity.\\\\
{\bf Acknowledgements}\\
This work is supported by the National Natural Science Foundation of
China (Grant No. 12375043).

\end{document}